\documentclass[twocolumn, preprintnumbers, amsmath, amssymb,prl,showpacs,longbibliography]{revtex4-1}

\usepackage[pagewise, modulo, displaymath, mathlines]{lineno}		

\usepackage{graphicx}  
\usepackage{dcolumn}   
\usepackage{bm}        
\usepackage{float}
\usepackage{natbib}
\usepackage{amssymb}
\usepackage{color}

\newcommand{\now}[1]{\textcolor{black}{#1}}

\begin{document}

\setlength\linenumbersep{8 mm}	
\title{Interference of single photons emitted by\\entangled atoms in free space}
\author{G. Araneda$^{1}$}
\email{gabriel.araneda-machuca@uibk.ac.at}
\author{D. B. Higginbottom$^{1,2}$}
\author{L. Slodi\v{c}ka$^{3}$}
\author{Y. Colombe$^1$}
\email{yves.colombe@uibk.ac.at}
\author{R. Blatt$^{1,4}$}
\affiliation{$^1$Institut f\"{u}r Experimentalphysik, Universit\"{a}t Innsbruck, Technikerstra\ss e 25, 6020 Innsbruck, Austria}
\affiliation{$^2$Centre for Quantum Computation and Communication Technology, Research School of Physics and Engineering, The Australian National University, Canberra ACT 2601, Australia.}
\affiliation{$^3$Department of Optics Palack\'{y} University, 17. Listopadu 12, 77146 Olomouc, Czech Republic}
\affiliation{$^4$Institut f\"{u}r Quantenoptik und Quanteninformation, \"{O}sterreichische Akademie der Wissenschaften, Technikerstra\ss e 21a, 6020 Innsbruck, Austria}

\begin{abstract}
The generation and manipulation of entanglement between isolated particles has precipitated rapid progress in quantum information processing. Entanglement is also known to play an essential role in the optical properties of atomic ensembles, but fundamental effects in the controlled emission and absorption from small, well-defined numbers of entangled emitters in free space have remained unobserved. Here we present the control of the \now{emission rate} of a single photon from a pair of distant, entangled atoms into a free-space optical mode. Changing the length of the optical path connecting the atoms modulates the \now{single photon emission rate in the selected mode} with a visibility \now{$V = 0.27 \pm 0.03$} determined by the degree of entanglement shared between the atoms, corresponding directly to the concurrence \now{ $\mathcal{C_{\rho}}= 0.31 \pm 0.10$} of the prepared state. This scheme, together with population measurements, provides a fully optical determination of the amount of entanglement. Furthermore, large sensitivity of the interference phase evolution points to applications of the presented scheme in high-precision gradient sensing.

\end{abstract}

\maketitle

Collective emission and absorption properties of entangled emitters have been extensively studied in ensembles of neutral atoms \cite{Schmied2016, McConnell2015, Scully2009, Haas2014a}, where in general, the number of emitters fluctuates and the control of the quantum state of each emitter is challenging.
To observe the coherent interaction of light with a definite number of entangled emitters it is necessary to achieve simultaneously sub-wavelength emitter localization, high-fidelity entanglement generation and atom-light coupling strong enough to detect an optical signal at very low photon flux. Experimental systems motivated by quantum computation have enabled excellent control over position and entanglement with well-defined and steadily increasing numbers of particles \cite{tan2015multi, ballance2015hybrid, monz201114}. Meanwhile, the pursuit of on-demand single photons, strong atom-light coupling, and fast quantum state readout has advanced the collection efficiency of light from single emitters \cite{Maiwald2012a, Streed2012, shu2011efficient, lodahl2015interfacing}. Together these developments enable the investigation of entanglement in collective atom-light interactions in the few-atom limit \cite{mlynek2014observation, casabone2015enhanced}.

\now{Previous experiments have shown free-space or cavity-mediated interference from atoms in separable states \cite{beugnon2006quantum,neuzner2016interference,Moehring2007}, where each atom emits a photon independently, or demonstrated enhancement and suppression of single-photon emission in a cavity mode from an entangled state of two atoms \cite{casabone2015enhanced}. In this Letter we present the first observation of interference in the spontaneous emission of a single photon into free space, emitted jointly by a pair of atoms prepared in a well-characterized entangled state.} The atoms are trapped adjacently but share a common optical mode in which their effective optical separation is $d \simeq 60$\,cm. The interference we observe in the single-photon emission probability arises solely from the entanglement present in the two-atom state \cite{wiegner2011a}. With this arrangement we directly observe both enhancement and inhibition of single-photon emission from the atoms by controlling the optical distance between them in the common mode. This distance can be set dynamically during an experimental sequence, with sub-wavelength precision and over time scales well below the lifetime of the entangled state.

\begin{figure*}[t]
\centerline{\includegraphics[width=0.90\textwidth]{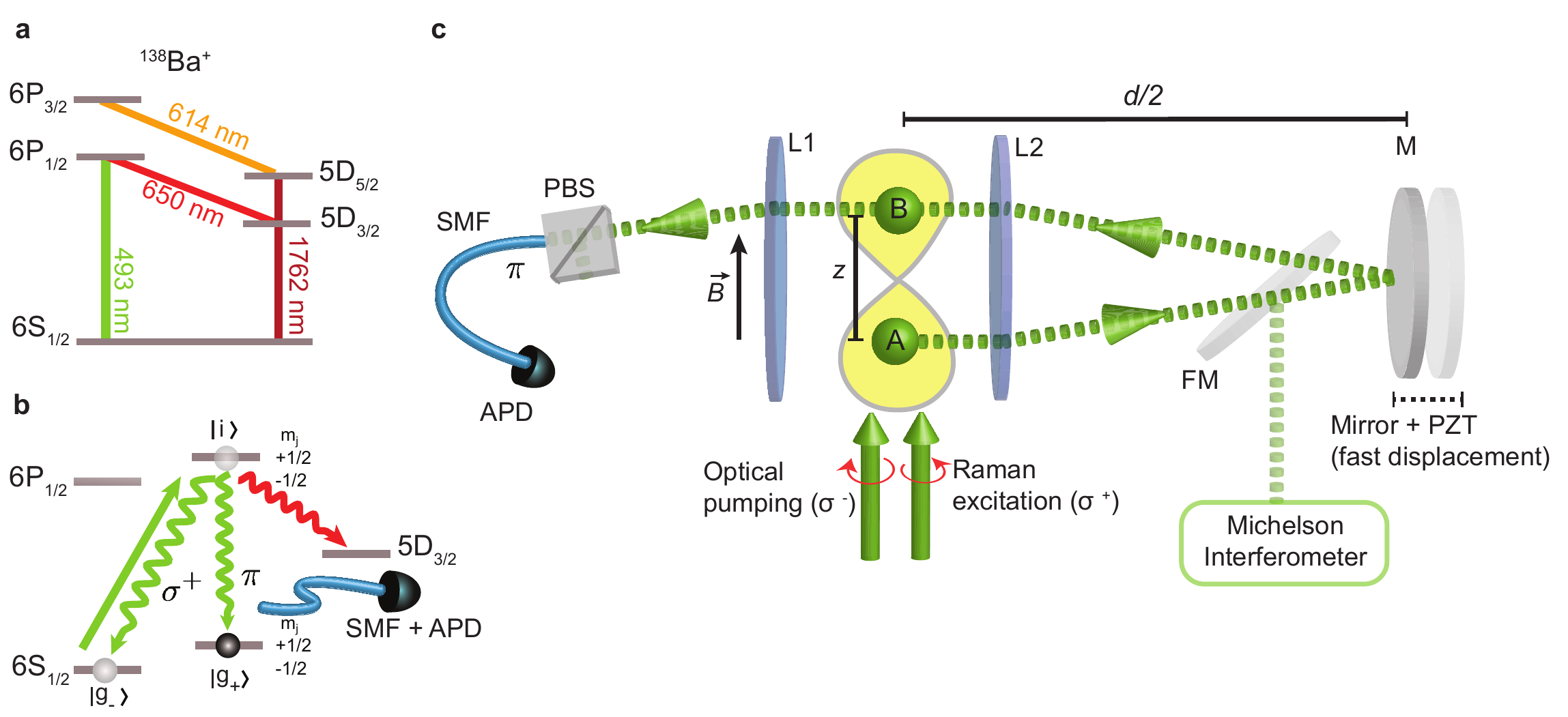}}
\caption{ (a), Electronic levels of $^{138}$Ba$^+$. (b), Excitation of the short-lived intermediate state $|\text{i}\rangle$ by a $\sigma^+$ beam at $493$\,nm. Detecting a $\pi$-polarized herald photon during a weak excitation pulse prepares a two-atom entangled state. A further, stronger excitation scatters a $\pi$-polarized witness photon from the entangled state in the common mode with a probability that depends on the phase difference $\Delta \phi$. (c), Two $^{138}$Ba$^+$ ions, A and B, are trapped and cooled in a linear Paul trap, separated by a distance $z \simeq 5.2\,\mu$m. The radiation fields are collimated using two identical in-vacuum high numerical aperture lenses (L1 and L2). A mirror (M) at a distance $d/2 \simeq 30\,$cm superimposes emissions from the ions so that they are coupled to a common spatial mode. The mirror is mounted on piezo transducers (PZT) that allow fast sub-wavelength control of the atom-atom distance in the common optical mode. A polarizing beam splitter (PBS) selects $\pi$-polarized photons, which are coupled to a single-mode fiber (SMF) and detected by an avalanche photodiode (APD). A Michelson interferometer enabled by a flip mirror (FM) is used to calibrate the PZT-mirror voltage-displacement response.}
\label{setup_fig_full}
\end{figure*}

\now{In order to study the optical properties of entangled particles, we first entangle two atoms. Entanglement can be achieved, e.g., with schemes relying on motional and internal atomic states \cite{cirac1995quantum,schmidt2003realization,sorensen1999quantum}, or through the coincident detection of two photons \cite{Moehring2007}. Here we use a scheme requiring only the detection of a single photon, proposed by Cabrillo \textit{et al.} \cite{cabrillo1999} and demonstrated in \cite{Slodicka2013}.} Two Doppler-cooled $^{138}$Ba$^+$ ions, with electronic structure shown in Fig.~\ref{setup_fig_full}(a), initially prepared in an electronic ground state $|\text{g}_-\rangle$ = $|6\text{S}_\frac{1}{2}, m_j = -\frac{1}{2}\rangle$ are excited to the state $|\text{g}_+\rangle$ = $|6\text{S}_\frac{1}{2}, m_j = +\frac{1}{2}\rangle$ with single-atom transition probability $p_{\text{e}} = 6\pm 1\,\%$ through the spontaneous Raman process $|\text{g}_-\rangle\rightarrow|\text{i}\rangle\rightarrow|\text{g}_+\rangle$ by weak laser excitation to the intermediate, short-lived excited state $|\text{i}\rangle = |6\text{P}_\frac{1}{2}, m_j = +\frac{1}{2}\rangle$, see Fig.~\ref{setup_fig_full}(b). A magnetic field $\vec{B}$ is applied along the trap axis, defining the quantization axis. Part of the emission of the atoms is coupled to a $\pi$-polarized, wavelength $\lambda = 493\,$nm common mode by using in-vacuum lenses (L1 and L2) and the distant mirror (M), see Fig.~\ref{setup_fig_full}(c). The emission is coupled into a single-mode fiber (SMF) defining the spatial mode, together with a polarizing beam splitter (PBS) defining its polarization. Detection of a single `herald' photon in this common mode projects the atoms onto the entangled state \cite{Slodicka2013}
\begin{equation}
|\psi\rangle = \frac{1}{\sqrt{2}}(|\text{g}_+,\text{g}_-\rangle + e^{i\phi}|\text{g}_-,\text{g}_+\rangle ),
\label{entangled_state}
\end{equation}
which belongs to a decoherence-free subspace \cite{haffner2005robust}. The probability of generating this state is $2p_\text{e}(1-p_\text{e})\eta$, where $\eta$ is the overall detection efficiency of a single photon. The phase $\phi$ of the entangled state is given by
\begin{equation}
\phi = (\phi_{\text{L}_\text{B}}-\phi_{\text{L}_\text{A}})+(\phi_{\text{D}_\text{B}}-\phi_{\text{D}_\text{A}}),
\end{equation}
where $\phi_{\text{L}_{\text{B}}}-\phi_{\text{L}_{\text{A}}} = k z$ is the phase difference of the exciting laser field at the positions of the atoms A and B, $\phi_{\text{D}_{\text{B}}}-\phi_{\text{D}_{\text{A}}} = -kd$ is the phase difference associated with the path followed by a photon emitted by atom A or B on its way to the detector, and $k = 2\pi/\lambda$.

We estimate a fidelity $F = 0.65 \pm 0.02$ between the experimentally generated state and the ideal entangled state $|\psi\rangle$ by measuring parity oscillations \cite{sackett2000experimental}. The spatial, spectral and polarization indistinguishability of the detected photons is characterized by the value of the second-order correlation function at zero delay $g^{(2)}(t = 0) = 0.99 \pm 0.06$, close to perfect indistinguishability, so that the fidelity is limited by other factors, \now{primarily atomic motion} (for details on entanglement generation and characterization see Supplementary Information sections 1 and 2). The herald photon detection rate, which corresponds to the rate of entanglement generation, is 5.02\,s$^{-1}$.

If we further excite the entangled atom pair with the $\sigma^+$-polarized beam, only the component of the entangled state in $|\text{g}_-\rangle$ may absorb a photon and spontaneously decay to $|\text{g}_+\rangle$, emitting a second photon. The state of the joint atom-photon system after completing this second Raman process is
\begin{equation}
|\psi'\rangle = \frac{1}{\sqrt{2}}(|0,1\rangle+e^{i(\phi_{\text{L}_\text{A}}-\phi_{\text{L}_\text{B}})}e^{i\phi}|1,0\rangle)|\text{g}_+,\text{g}_+\rangle.
\label{second_state_disting}
\end{equation}
The field states $|1,0\rangle$ and $|0,1\rangle$ correspond to the emission of a single photon from atom A or B. The detection of a `witness' photon in the common mode projects the atoms onto the unnormalized state
\begin{equation}
|\psi'_p\rangle = \frac{1}{\sqrt{2}}(1+e^{i(\phi_{\text{L}_\text{A}}-\phi_{\text{L}_\text{B}}+\phi'_{\text{D}_\text{A}}-\phi'_{\text{D}_\text{B}})}e^{i\phi})|\text{g}_+,\text{g}_+\rangle,
\label{second_state}
\end{equation}
where $\phi'_{\text{D}_\text{A}}-\phi'_{\text{D}_\text{B}} = k d'$ is the optical phase difference in the common mode when the witness photon is detected. The probability of detecting a single witness photon is
\begin{equation}
P \propto \left| \langle \psi'_p|\psi'_p\rangle \right|^2 = 1 + \cos(\phi - \phi'),
\label{intensity}
\end{equation}
where $\phi' = (\phi_{\text{L}_\text{B}}-\phi_{\text{L}_\text{A}})+(\phi'_{\text{D}_\text{B}}-\phi'_{\text{D}_\text{A}})$. The witness photon detection probability is modulated by the phase difference $\Delta\phi = \phi- \phi'$ between the herald and witness detection events. This effect is a consequence of entanglement between the two emitters and corresponds to enhancement or inhibition of the emission probability in the common mode due to single-photon path interference \cite{wiegner2011a}.

\begin{figure}[t!]
\centerline{\includegraphics[width=0.95\columnwidth]{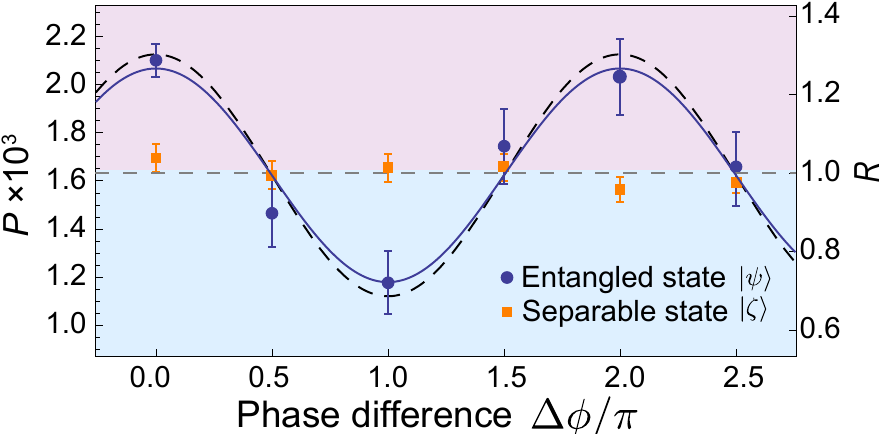}}
\caption{Absolute ($P$) and relative ($R$) witness photon probability from the atom pair as a function of the phase difference $\Delta \phi$ for the entangled state $|\psi\rangle$ (blue circles) and the separable state $|\zeta\rangle$ (orange squares). Error bars correspond to the Poissonian error from photon counting. The blue solid curve is the amplitude and offset fitted scattering model for the entangled state, while the black dashed curve shows the expected probability from the independent estimation of its concurrence. The gray dashed line is the average of the fitted curve. The witness photon detection probability for $|\psi\rangle$ is maximally enhanced at $\Delta \phi = 0$, $2\pi$ and maximally suppressed at $\Delta \phi = \pi$. The photon detection probability for $|\zeta\rangle$ is constant within the measurement uncertainty.}
\label{mirror_delay}
\end{figure}

We vary the phase difference $\Delta \phi$ by rapidly changing the position of the distant mirror from $d/2$ to $d'/2$ between the two detection events within a time $\tau \leq 220\,\mu$s, while maintaining a fixed interatomic distance $z$, so that $\Delta\phi = k(d'-d)$ (Supplementary Information section 3). This allows precise phase control in short times without motional excitation of the ions. In order to efficiently generate a witness photon from the entangled state a laser pulse, stronger than the pulse used to create entanglement, is used, with a $|\text{g}_-\rangle\rightarrow|\text{g}_+\rangle$ transition probability $p_\text{w} =80 \pm 2\,\%$ for a single atom. The full sequence is described in Supplementary Information section 4. We measure the witness photon detection probability $P$ by counting the number of witness photons per herald photon for a given phase change $\Delta \phi$. The results of the measurement, together with the amplitude- and offset-fitted model from Eq.~\eqref{intensity}, are plotted in Fig.~\ref{mirror_delay}. The maximum and minimum measured probabilities are $P(\Delta \phi = 0) = (2.10\, \pm\, 0.07) \times 10^{-3}$ and $P(\Delta \phi = \pi) = (1.17\, \pm\, 0.12) \times 10^{-3}$.
It has been established theoretically, but not previously observed, that the visibility of the interference fringes in radiation from a pair of two-level emitters with a single excitation should be equal to the concurrence of the bipartite quantum state \cite{suzuki2010entanglement,scholak2008entanglement}, see Supplementary Information section 5. The visibility $V$ obtained by the fitted model implies then a concurrence of the entangled atom pair $\mathcal{C_\text{wit}} = V = 0.27 \pm 0.03$, in agreement with the concurrence of the bipartite density matrix inferred from parity measurements $\mathcal{C(\rho_\text{par})} = 0.31 \pm 0.10$. \now{The fidelity of the entangled state, and therefore the visibility of the fringes, is limited by the motion of the atoms (Supplementary Information section 2).}

For comparison we measure the photon detection rate for a separable two-atom state $|\zeta\rangle = |\text{g}_+,\text{g}_-\rangle$. This state is prepared using a combination of optical pumping in the $|\text{g}_-\rangle$ state, a $1.76\,\mu$m addressed shelving pulse on the $|\text{g}_-\rangle \leftrightarrow |5\text{D}_{\frac{5}{2}},m_j = -\frac{5}{2}\rangle$ transition on atom B, a global RF $\pi$-pulse between the $|\text{g}_-\rangle$ and $|\text{g}_+\rangle$ states of both ions, before un-shelving atom B. We trigger the emission of a witness photon from this state by using an excitation beam with the same parameters used for emitting a photon from the state $|\psi\rangle$. The witness photon detection probability for $|\zeta\rangle$, which is the mean number of photons detected per prepared state, is shown in Fig.~\ref{mirror_delay} (orange squares) as a function of the optical phase $\Delta\phi$. In contrast to the entangled state $|\psi\rangle$, the photon probability $P_{|\zeta\rangle}$ for the separable state $|\zeta\rangle$ is independent of the phase difference, with an average detection probability $P_{\text{sep}} = (1.63 \,\pm\,0.05)\times 10^{-3}$. The observed visibility for this state is $\approx 0$, in agreement with the expected vanishing concurrence of the state. We define the relative probabilities $R = P(\Delta \phi) / P_\text{sep}$, so that $R > 1$ ($R < 1$) represents enhanced (suppressed) detection probability relative to this separable state. The relative scale is shown in Fig.~\ref{mirror_delay}, right vertical axis. Because states $|\psi\rangle$ and $|\zeta\rangle$ both contain a single atom in $|\text{g}_-\rangle$, we expect the mean detection probabilities from each state to be equal (Supplementary Information section 5). The mean of the fitted interference curve for $R_{|\psi\rangle}$ in Fig.~\ref{mirror_delay} is $0.99 \pm 0.08$, in close agreement with the mean of $R_{|\zeta\rangle}$. The case of a more general separable state is studied in Supplementary Information section 5.


\begin{figure}[t!]
\centerline{\includegraphics[width=0.95\columnwidth]{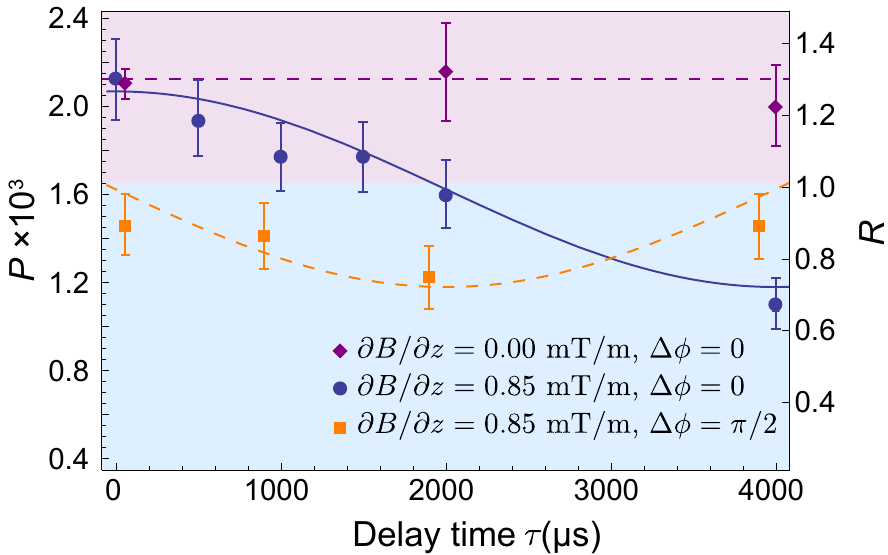}}
\caption{ Absolute ($P$) and relative ($R$) witness photon probability when $\Delta \phi = 0 $ (blue circles) and $\Delta \phi = \pi/2$ (orange squares) for delay times $\tau$ in the presence of a magnetic field gradient compared to the same measure in the absence of a magnetic field gradient (purple diamonds) with $\Delta \phi = 0 $. The blue curve is amplitude and period fitted to the measured data for $\Delta \phi = 0$ (blue points), while the orange dashed curve is a $\pi/2$ phase shifted version of this fit, showing with the $\Delta\phi = \pi/2$ measured data. The dashed purple line shows constant enhancement in the absence of magnetic field gradient. }
\label{delay}
\end{figure}

The witness photon detection probability $P$ is also sensitive to any phase accumulated by the entangled state between the emission of herald and witness photons. For example, the presence of a static magnetic field gradient between the ions induces a linear increase of the entangled state phase $\phi$ \cite{schmidt2012entangled,ruster2017entanglement}. We measure the magnetic field gradient due to an external permanent magnet by recording the evolution of the witness photon emission probability with a variable delay between the herald and witness photons. The results are shown in Fig.~\ref{delay} for mirror displacements corresponding to $\Delta \phi = 0$ (blue) and $\Delta \phi = \pi/2$ (orange). Oscillations observed in both signals correspond to a linear evolution of the entanglement phase with a $\pi/2$ phase shift respect to each other. The period of the oscillation obtained from the fit (blue curve) of the data for $\Delta \phi = 0$ is 8.0 $\pm$ 0.5 ms. The orange curve is the same fit shifted by $\pi/2$, and shows agreement with the measured data for $\Delta\phi = \pi/2$. The period of the oscillation implies a magnetic field gradient of 0.85 $\pm$ 0.05 mT/m along the ion crystal. The same measurement is taken in absence of magnetic field gradient for $\Delta\phi = 0$ (Fig.~\ref{delay}, purple diamonds), where the signal remains constant since dephasing effects are negligible on the measured time scale. The coherence time of the entangled state in the decoherence-free subspace is expected to be orders of magnitude longer than the time scale of the shown phase evolution \cite{haffner2005robust}.

\now{While super- and subradiance, i.e., modifications of the global spontaneous emission rate, have been previously observed with a pair of trapped atoms \cite{devoe1996observation}, here we have shown the control of the single-photon emission rate into a selected free-space optical mode.} 
This opens the way to experimental studies of optical properties of entangled states of a few well-controlled emitters \cite{suzuki2010entanglement, ficek2002entangled, wiegner2011a, navarrete2011simulating} such as trapped ions as presented here, or neutral atoms in more versatile trapping configurations \cite{barredo2016atom}. The free-space configuration imposes no fundamental constraints on the direction of emission, the number of entangled particles, or their mutual distance. The phases of the emitters in the detected optical mode can be tuned arbitrarily, allowing the observation of the complete interference signal; this is in contrast to Fabry-Perot cavity systems \cite{mlynek2014observation, casabone2015enhanced} in which the relative phase of the emitters is restricted to 0 or $\pi$ by the cavity mode. The agreement between measurements of entanglement using single-photon interference visibility and independent measurements of internal states populations confirms that the observed interference can be employed for the estimation of entanglement between distant particles, without any requirement on coherent control over internal atomic states \cite{jakob2010quantitative, suzuki2010entanglement, ficek2002entangled}. Such a scheme is also applicable to quantum objects with different internal structures and emission spectra \cite{vittorini2014entanglement}, and should allow the observation of entanglement between non-identical atoms or general disparate quantum emitters \cite{tan2015multi, kurizki2015quantum, ficek2002entangled, wiegner2011IOP}. 

The spatial differences of various environmental factors are mapped directly into interference of photons from entangled particles through the evolution of the phase of the entangled state, as shown here for the case of magnetic field gradients. This points to potential applications in quantum metrology \cite{giovannetti2011advances}; together with all-optical preparation of distant entangled states \cite{Slodicka2013, Moehring2007} and recent developments in the stabilization of large fiber networks \cite{predehl2012920}, the technique presented here may enable the development of ultra-sensitive optical gradiometers \cite{yurtsever2003interferometry}.

\vspace{10pt}
We acknowledge the technical support of Muir Kumph, Matthias Brandl, Nicolas Chauvet, Bernardo Casabone, and Bryan Luu. This work has been financially supported by the Austrian Science Fund (FWF) through projects P23022 (SINPHONIA) and F4001 (SFB FoQuS), by the European Research Council (ERC) project CRYTERION \#227959, and the Institut f\"ur Quanteninformation GmbH. L.S. acknowledges financial support from the Grant Agency of the Czech Republic, Grant No. GB14-36681G. D.B.H. acknowledges support from the Australian Research Council (ARC) (CE1101027).


\begin{thebibliography}{39}%
\makeatletter
\providecommand \@ifxundefined [1]{%
 \@ifx{#1\undefined}
}%
\providecommand \@ifnum [1]{%
 \ifnum #1\expandafter \@firstoftwo
 \else \expandafter \@secondoftwo
 \fi
}%
\providecommand \@ifx [1]{%
 \ifx #1\expandafter \@firstoftwo
 \else \expandafter \@secondoftwo
 \fi
}%
\providecommand \natexlab [1]{#1}%
\providecommand \enquote  [1]{``#1''}%
\providecommand \bibnamefont  [1]{#1}%
\providecommand \bibfnamefont [1]{#1}%
\providecommand \citenamefont [1]{#1}%
\providecommand \href@noop [0]{\@secondoftwo}%
\providecommand \href [0]{\begingroup \@sanitize@url \@href}%
\providecommand \@href[1]{\@@startlink{#1}\@@href}%
\providecommand \@@href[1]{\endgroup#1\@@endlink}%
\providecommand \@sanitize@url [0]{\catcode `\\12\catcode `\$12\catcode
  `\&12\catcode `\#12\catcode `\^12\catcode `\_12\catcode `\%12\relax}%
\providecommand \@@startlink[1]{}%
\providecommand \@@endlink[0]{}%
\providecommand \url  [0]{\begingroup\@sanitize@url \@url }%
\providecommand \@url [1]{\endgroup\@href {#1}{\urlprefix }}%
\providecommand \urlprefix  [0]{URL }%
\providecommand \Eprint [0]{\href }%
\providecommand \doibase [0]{http://dx.doi.org/}%
\providecommand \selectlanguage [0]{\@gobble}%
\providecommand \bibinfo  [0]{\@secondoftwo}%
\providecommand \bibfield  [0]{\@secondoftwo}%
\providecommand \translation [1]{[#1]}%
\providecommand \BibitemOpen [0]{}%
\providecommand \bibitemStop [0]{}%
\providecommand \bibitemNoStop [0]{.\EOS\space}%
\providecommand \EOS [0]{\spacefactor3000\relax}%
\providecommand \BibitemShut  [1]{\csname bibitem#1\endcsname}%
\let\auto@bib@innerbib\@empty
\bibitem [{\citenamefont {Schmied}\ \emph {et~al.}(2016)\citenamefont
  {Schmied}, \citenamefont {Bancal}, \citenamefont {Allard}, \citenamefont
  {Fadel}, \citenamefont {Scarani}, \citenamefont {Treutlein},\ and\
  \citenamefont {Sangouard}}]{Schmied2016}%
  \BibitemOpen
  \bibfield  {author} {\bibinfo {author} {\bibfnamefont {R.}~\bibnamefont
  {Schmied}}, \bibinfo {author} {\bibfnamefont {J.-D.}\ \bibnamefont {Bancal}},
  \bibinfo {author} {\bibfnamefont {B.}~\bibnamefont {Allard}}, \bibinfo
  {author} {\bibfnamefont {M.}~\bibnamefont {Fadel}}, \bibinfo {author}
  {\bibfnamefont {V.}~\bibnamefont {Scarani}}, \bibinfo {author} {\bibfnamefont
  {P.}~\bibnamefont {Treutlein}}, \ and\ \bibinfo {author} {\bibfnamefont
  {N.}~\bibnamefont {Sangouard}},\ }\href@noop {} {\bibfield  {journal}
  {\bibinfo  {journal} {Science}\ }\textbf {\bibinfo {volume} {352}},\ \bibinfo
  {pages} {441} (\bibinfo {year} {2016})}\BibitemShut {NoStop}%
\bibitem [{\citenamefont {McConnell}\ \emph {et~al.}(2015)\citenamefont
  {McConnell}, \citenamefont {Zhang}, \citenamefont {Hu}, \citenamefont
  {{\'{C}}uk},\ and\ \citenamefont {Vuleti{\'{c}}}}]{McConnell2015}%
  \BibitemOpen
  \bibfield  {author} {\bibinfo {author} {\bibfnamefont {R.}~\bibnamefont
  {McConnell}}, \bibinfo {author} {\bibfnamefont {H.}~\bibnamefont {Zhang}},
  \bibinfo {author} {\bibfnamefont {J.}~\bibnamefont {Hu}}, \bibinfo {author}
  {\bibfnamefont {S.}~\bibnamefont {{\'{C}}uk}}, \ and\ \bibinfo {author}
  {\bibfnamefont {V.}~\bibnamefont {Vuleti{\'{c}}}},\ }\href@noop {} {\bibfield
   {journal} {\bibinfo  {journal} {Nature (London)}\ }\textbf {\bibinfo
  {volume} {519}},\ \bibinfo {pages} {439} (\bibinfo {year}
  {2015})}\BibitemShut {NoStop}%
\bibitem [{\citenamefont {Scully}\ and\ \citenamefont
  {Svidzinsky}(2009)}]{Scully2009}%
  \BibitemOpen
  \bibfield  {author} {\bibinfo {author} {\bibfnamefont {M.~O.}\ \bibnamefont
  {Scully}}\ and\ \bibinfo {author} {\bibfnamefont {A.~A.}\ \bibnamefont
  {Svidzinsky}},\ }\href@noop {} {\bibfield  {journal} {\bibinfo  {journal}
  {Science}\ }\textbf {\bibinfo {volume} {325}},\ \bibinfo {pages} {1510}
  (\bibinfo {year} {2009})}\BibitemShut {NoStop}%
\bibitem [{\citenamefont {Haas}\ \emph {et~al.}(2014)\citenamefont {Haas},
  \citenamefont {Volz}, \citenamefont {Gehr}, \citenamefont {Reichel},\ and\
  \citenamefont {Est{\`{e}}ve}}]{Haas2014a}%
  \BibitemOpen
  \bibfield  {author} {\bibinfo {author} {\bibfnamefont {F.}~\bibnamefont
  {Haas}}, \bibinfo {author} {\bibfnamefont {J.}~\bibnamefont {Volz}}, \bibinfo
  {author} {\bibfnamefont {R.}~\bibnamefont {Gehr}}, \bibinfo {author}
  {\bibfnamefont {J.}~\bibnamefont {Reichel}}, \ and\ \bibinfo {author}
  {\bibfnamefont {J.}~\bibnamefont {Est{\`{e}}ve}},\ }\href@noop {} {\bibfield
  {journal} {\bibinfo  {journal} {Science}\ }\textbf {\bibinfo {volume}
  {344}},\ \bibinfo {pages} {180} (\bibinfo {year} {2014})}\BibitemShut
  {NoStop}%
\bibitem [{\citenamefont {Tan}\ \emph {et~al.}(2015)\citenamefont {Tan},
  \citenamefont {Gaebler}, \citenamefont {Lin}, \citenamefont {Wan},
  \citenamefont {Bowler}, \citenamefont {Leibfried},\ and\ \citenamefont
  {Wineland}}]{tan2015multi}%
  \BibitemOpen
  \bibfield  {author} {\bibinfo {author} {\bibfnamefont {T.~R.}\ \bibnamefont
  {Tan}}, \bibinfo {author} {\bibfnamefont {J.~P.}\ \bibnamefont {Gaebler}},
  \bibinfo {author} {\bibfnamefont {Y.}~\bibnamefont {Lin}}, \bibinfo {author}
  {\bibfnamefont {Y.}~\bibnamefont {Wan}}, \bibinfo {author} {\bibfnamefont
  {R.}~\bibnamefont {Bowler}}, \bibinfo {author} {\bibfnamefont
  {D.}~\bibnamefont {Leibfried}}, \ and\ \bibinfo {author} {\bibfnamefont
  {D.~J.}\ \bibnamefont {Wineland}},\ }\href@noop {} {\bibfield  {journal}
  {\bibinfo  {journal} {Nature (London)}\ }\textbf {\bibinfo {volume} {528}},\
  \bibinfo {pages} {380} (\bibinfo {year} {2015})}\BibitemShut {NoStop}%
\bibitem [{\citenamefont {Ballance}\ \emph {et~al.}(2015)\citenamefont
  {Ballance}, \citenamefont {Sch{\"a}fer}, \citenamefont {Home}, \citenamefont
  {Szwer}, \citenamefont {Webster}, \citenamefont {Allcock}, \citenamefont
  {Linke},\ and\ \citenamefont {Harty}}]{ballance2015hybrid}%
  \BibitemOpen
  \bibfield  {author} {\bibinfo {author} {\bibfnamefont {C.~J.}\ \bibnamefont
  {Ballance}}, \bibinfo {author} {\bibfnamefont {V.~M.}\ \bibnamefont
  {Sch{\"a}fer}}, \bibinfo {author} {\bibfnamefont {J.~P.}\ \bibnamefont
  {Home}}, \bibinfo {author} {\bibfnamefont {D.}~\bibnamefont {Szwer}},
  \bibinfo {author} {\bibfnamefont {S.~C.}\ \bibnamefont {Webster}}, \bibinfo
  {author} {\bibfnamefont {D.~T.~C.}\ \bibnamefont {Allcock}}, \bibinfo
  {author} {\bibfnamefont {N.~M.}\ \bibnamefont {Linke}}, \ and\ \bibinfo
  {author} {\bibfnamefont {T.}~\bibnamefont {Harty}},\ }\href@noop {}
  {\bibfield  {journal} {\bibinfo  {journal} {Nature (London)}\ }\textbf
  {\bibinfo {volume} {528}},\ \bibinfo {pages} {384} (\bibinfo {year}
  {2015})}\BibitemShut {NoStop}%
\bibitem [{\citenamefont {Monz}\ \emph {et~al.}(2011)\citenamefont {Monz},
  \citenamefont {Schindler}, \citenamefont {Barreiro}, \citenamefont {Chwalla},
  \citenamefont {Nigg}, \citenamefont {Coish}, \citenamefont {Harlander},
  \citenamefont {H{\"a}nsel}, \citenamefont {Hennrich},\ and\ \citenamefont
  {Blatt}}]{monz201114}%
  \BibitemOpen
  \bibfield  {author} {\bibinfo {author} {\bibfnamefont {T.}~\bibnamefont
  {Monz}}, \bibinfo {author} {\bibfnamefont {P.}~\bibnamefont {Schindler}},
  \bibinfo {author} {\bibfnamefont {J.~T.}\ \bibnamefont {Barreiro}}, \bibinfo
  {author} {\bibfnamefont {M.}~\bibnamefont {Chwalla}}, \bibinfo {author}
  {\bibfnamefont {D.}~\bibnamefont {Nigg}}, \bibinfo {author} {\bibfnamefont
  {W.~A.}\ \bibnamefont {Coish}}, \bibinfo {author} {\bibfnamefont
  {M.}~\bibnamefont {Harlander}}, \bibinfo {author} {\bibfnamefont
  {W.}~\bibnamefont {H{\"a}nsel}}, \bibinfo {author} {\bibfnamefont
  {M.}~\bibnamefont {Hennrich}}, \ and\ \bibinfo {author} {\bibfnamefont
  {R.}~\bibnamefont {Blatt}},\ }\href@noop {} {\bibfield  {journal} {\bibinfo
  {journal} {Phys. Rev. Lett.}\ }\textbf {\bibinfo {volume} {106}},\ \bibinfo
  {pages} {130506} (\bibinfo {year} {2011})}\BibitemShut {NoStop}%
\bibitem [{\citenamefont {Maiwald}\ \emph {et~al.}(2012)\citenamefont
  {Maiwald}, \citenamefont {Golla}, \citenamefont {Fischer}, \citenamefont
  {Bader}, \citenamefont {Heugel}, \citenamefont {Chalopin}, \citenamefont
  {Sondermann},\ and\ \citenamefont {Leuchs}}]{Maiwald2012a}%
  \BibitemOpen
  \bibfield  {author} {\bibinfo {author} {\bibfnamefont {R.}~\bibnamefont
  {Maiwald}}, \bibinfo {author} {\bibfnamefont {A.}~\bibnamefont {Golla}},
  \bibinfo {author} {\bibfnamefont {M.}~\bibnamefont {Fischer}}, \bibinfo
  {author} {\bibfnamefont {M.}~\bibnamefont {Bader}}, \bibinfo {author}
  {\bibfnamefont {S.}~\bibnamefont {Heugel}}, \bibinfo {author} {\bibfnamefont
  {B.}~\bibnamefont {Chalopin}}, \bibinfo {author} {\bibfnamefont
  {M.}~\bibnamefont {Sondermann}}, \ and\ \bibinfo {author} {\bibfnamefont
  {G.}~\bibnamefont {Leuchs}},\ }\href@noop {} {\bibfield  {journal} {\bibinfo
  {journal} {Phys. Rev. A}\ }\textbf {\bibinfo {volume} {86}},\ \bibinfo
  {pages} {043431} (\bibinfo {year} {2012})}\BibitemShut {NoStop}%
\bibitem [{\citenamefont {Streed}\ \emph {et~al.}(2012)\citenamefont {Streed},
  \citenamefont {Jechow}, \citenamefont {Norton},\ and\ \citenamefont
  {Kielpinski}}]{Streed2012}%
  \BibitemOpen
  \bibfield  {author} {\bibinfo {author} {\bibfnamefont {E.~W.}\ \bibnamefont
  {Streed}}, \bibinfo {author} {\bibfnamefont {A.}~\bibnamefont {Jechow}},
  \bibinfo {author} {\bibfnamefont {B.~G.}\ \bibnamefont {Norton}}, \ and\
  \bibinfo {author} {\bibfnamefont {D.}~\bibnamefont {Kielpinski}},\
  }\href@noop {} {\bibfield  {journal} {\bibinfo  {journal} {Nat. Commun.}\
  }\textbf {\bibinfo {volume} {3}},\ \bibinfo {pages} {933} (\bibinfo {year}
  {2012})}\BibitemShut {NoStop}%
\bibitem [{\citenamefont {Shu}\ \emph {et~al.}(2011)\citenamefont {Shu},
  \citenamefont {Chou}, \citenamefont {Kurz}, \citenamefont {Dietrich},\ and\
  \citenamefont {Blinov}}]{shu2011efficient}%
  \BibitemOpen
  \bibfield  {author} {\bibinfo {author} {\bibfnamefont {G.}~\bibnamefont
  {Shu}}, \bibinfo {author} {\bibfnamefont {C.}~\bibnamefont {Chou}}, \bibinfo
  {author} {\bibfnamefont {N.}~\bibnamefont {Kurz}}, \bibinfo {author}
  {\bibfnamefont {M.~R.}\ \bibnamefont {Dietrich}}, \ and\ \bibinfo {author}
  {\bibfnamefont {B.~B.}\ \bibnamefont {Blinov}},\ }\href@noop {} {\bibfield
  {journal} {\bibinfo  {journal} {J. Opt. Soc. Am. B}\ }\textbf {\bibinfo
  {volume} {28}},\ \bibinfo {pages} {2865} (\bibinfo {year}
  {2011})}\BibitemShut {NoStop}%
\bibitem [{\citenamefont {Lodahl}\ \emph {et~al.}(2015)\citenamefont {Lodahl},
  \citenamefont {Mahmoodian}, ,\ and\ \citenamefont
  {Stobbe}}]{lodahl2015interfacing}%
  \BibitemOpen
  \bibfield  {author} {\bibinfo {author} {\bibfnamefont {P.}~\bibnamefont
  {Lodahl}}, \bibinfo {author} {\bibfnamefont {S.}~\bibnamefont {Mahmoodian}},
  , \ and\ \bibinfo {author} {\bibfnamefont {S.}~\bibnamefont {Stobbe}},\
  }\href@noop {} {\bibfield  {journal} {\bibinfo  {journal} {Rev. Mod. Phys.}\
  }\textbf {\bibinfo {volume} {87}},\ \bibinfo {pages} {347} (\bibinfo {year}
  {2015})}\BibitemShut {NoStop}%
\bibitem [{\citenamefont {Mlynek}\ \emph {et~al.}(2014)\citenamefont {Mlynek},
  \citenamefont {Abdumalikov}, \citenamefont {Eichler},\ and\ \citenamefont
  {Wallraff}}]{mlynek2014observation}%
  \BibitemOpen
  \bibfield  {author} {\bibinfo {author} {\bibfnamefont {J.~A.}\ \bibnamefont
  {Mlynek}}, \bibinfo {author} {\bibfnamefont {A.~A.}\ \bibnamefont
  {Abdumalikov}}, \bibinfo {author} {\bibfnamefont {C.}~\bibnamefont
  {Eichler}}, \ and\ \bibinfo {author} {\bibfnamefont {A.}~\bibnamefont
  {Wallraff}},\ }\href@noop {} {\bibfield  {journal} {\bibinfo  {journal} {Nat.
  Commun.}\ }\textbf {\bibinfo {volume} {5}},\ \bibinfo {pages} {5186}
  (\bibinfo {year} {2014})}\BibitemShut {NoStop}%
\bibitem [{\citenamefont {Casabone}\ \emph {et~al.}(2015)\citenamefont
  {Casabone}, \citenamefont {Friebe}, \citenamefont {Brandst{\"a}tter},
  \citenamefont {Sch{\"u}ppert}, \citenamefont {Blatt},\ and\ \citenamefont
  {Northup}}]{casabone2015enhanced}%
  \BibitemOpen
  \bibfield  {author} {\bibinfo {author} {\bibfnamefont {B.}~\bibnamefont
  {Casabone}}, \bibinfo {author} {\bibfnamefont {K.}~\bibnamefont {Friebe}},
  \bibinfo {author} {\bibfnamefont {B.}~\bibnamefont {Brandst{\"a}tter}},
  \bibinfo {author} {\bibfnamefont {K.}~\bibnamefont {Sch{\"u}ppert}}, \bibinfo
  {author} {\bibfnamefont {R.}~\bibnamefont {Blatt}}, \ and\ \bibinfo {author}
  {\bibfnamefont {T.}~\bibnamefont {Northup}},\ }\href@noop {} {\bibfield
  {journal} {\bibinfo  {journal} {Phys. Rev. Lett.}\ }\textbf {\bibinfo
  {volume} {114}},\ \bibinfo {pages} {023602} (\bibinfo {year}
  {2015})}\BibitemShut {NoStop}%
\bibitem [{\citenamefont {Beugnon}\ \emph {et~al.}(2006)\citenamefont
  {Beugnon}, \citenamefont {Jones}, \citenamefont {Dingjan}, \citenamefont
  {Darqui{\'e}}, \citenamefont {Messin}, \citenamefont {Browaeys},\ and\
  \citenamefont {Grangier}}]{beugnon2006quantum}%
  \BibitemOpen
  \bibfield  {author} {\bibinfo {author} {\bibfnamefont {J.}~\bibnamefont
  {Beugnon}}, \bibinfo {author} {\bibfnamefont {M.}~\bibnamefont {Jones}},
  \bibinfo {author} {\bibfnamefont {J.}~\bibnamefont {Dingjan}}, \bibinfo
  {author} {\bibfnamefont {B.}~\bibnamefont {Darqui{\'e}}}, \bibinfo {author}
  {\bibfnamefont {G.}~\bibnamefont {Messin}}, \bibinfo {author} {\bibfnamefont
  {A.}~\bibnamefont {Browaeys}}, \ and\ \bibinfo {author} {\bibfnamefont
  {P.}~\bibnamefont {Grangier}},\ }\href@noop {} {\bibfield  {journal}
  {\bibinfo  {journal} {Nature}\ }\textbf {\bibinfo {volume} {440}},\ \bibinfo
  {pages} {779} (\bibinfo {year} {2006})}\BibitemShut {NoStop}%
\bibitem [{\citenamefont {Neuzner}\ \emph {et~al.}(2016)\citenamefont
  {Neuzner}, \citenamefont {K{\"o}rber}, \citenamefont {Morin}, \citenamefont
  {Ritter},\ and\ \citenamefont {Rempe}}]{neuzner2016interference}%
  \BibitemOpen
  \bibfield  {author} {\bibinfo {author} {\bibfnamefont {A.}~\bibnamefont
  {Neuzner}}, \bibinfo {author} {\bibfnamefont {M.}~\bibnamefont {K{\"o}rber}},
  \bibinfo {author} {\bibfnamefont {O.}~\bibnamefont {Morin}}, \bibinfo
  {author} {\bibfnamefont {S.}~\bibnamefont {Ritter}}, \ and\ \bibinfo {author}
  {\bibfnamefont {G.}~\bibnamefont {Rempe}},\ }\href@noop {} {\bibfield
  {journal} {\bibinfo  {journal} {Nature Photonics}\ }\textbf {\bibinfo
  {volume} {10}},\ \bibinfo {pages} {303} (\bibinfo {year} {2016})}\BibitemShut
  {NoStop}%
\bibitem [{\citenamefont {Moehring}\ \emph {et~al.}(2007)\citenamefont
  {Moehring}, \citenamefont {Maunz}, \citenamefont {Olmschenk}, \citenamefont
  {Younge}, \citenamefont {Matsukevich}, \citenamefont {Duan},\ and\
  \citenamefont {Monroe}}]{Moehring2007}%
  \BibitemOpen
  \bibfield  {author} {\bibinfo {author} {\bibfnamefont {D.~L.}\ \bibnamefont
  {Moehring}}, \bibinfo {author} {\bibfnamefont {P.}~\bibnamefont {Maunz}},
  \bibinfo {author} {\bibfnamefont {S.}~\bibnamefont {Olmschenk}}, \bibinfo
  {author} {\bibfnamefont {K.~C.}\ \bibnamefont {Younge}}, \bibinfo {author}
  {\bibfnamefont {D.~N.}\ \bibnamefont {Matsukevich}}, \bibinfo {author}
  {\bibfnamefont {L.-M.}\ \bibnamefont {Duan}}, \ and\ \bibinfo {author}
  {\bibfnamefont {C.}~\bibnamefont {Monroe}},\ }\href@noop {} {\bibfield
  {journal} {\bibinfo  {journal} {Nature (London)}\ }\textbf {\bibinfo {volume}
  {449}},\ \bibinfo {pages} {68} (\bibinfo {year} {2007})}\BibitemShut
  {NoStop}%
\bibitem [{\citenamefont {Wiegner}\ \emph
  {et~al.}(2011{\natexlab{a}})\citenamefont {Wiegner}, \citenamefont {von
  Zanthier},\ and\ \citenamefont {Agarwal}}]{wiegner2011a}%
  \BibitemOpen
  \bibfield  {author} {\bibinfo {author} {\bibfnamefont {R.}~\bibnamefont
  {Wiegner}}, \bibinfo {author} {\bibfnamefont {J.}~\bibnamefont {von
  Zanthier}}, \ and\ \bibinfo {author} {\bibfnamefont {G.~S.}\ \bibnamefont
  {Agarwal}},\ }\href@noop {} {\bibfield  {journal} {\bibinfo  {journal} {Phys.
  Rev. A.}\ }\textbf {\bibinfo {volume} {84}},\ \bibinfo {pages} {023805}
  (\bibinfo {year} {2011}{\natexlab{a}})}\BibitemShut {NoStop}%
\bibitem [{\citenamefont {Cirac}\ and\ \citenamefont
  {Zoller}(1995)}]{cirac1995quantum}%
  \BibitemOpen
  \bibfield  {author} {\bibinfo {author} {\bibfnamefont {J.~I.}\ \bibnamefont
  {Cirac}}\ and\ \bibinfo {author} {\bibfnamefont {P.}~\bibnamefont {Zoller}},\
  }\href@noop {} {\bibfield  {journal} {\bibinfo  {journal} {Phys. Re. Lett.}\
  }\textbf {\bibinfo {volume} {74}},\ \bibinfo {pages} {4091} (\bibinfo {year}
  {1995})}\BibitemShut {NoStop}%
\bibitem [{\citenamefont {Schmidt-Kaler}\ \emph {et~al.}(2003)\citenamefont
  {Schmidt-Kaler}, \citenamefont {H{\"a}ffner}, \citenamefont {Riebe},
  \citenamefont {Gulde}, \citenamefont {Lancaster}, \citenamefont {Deuschle},
  \citenamefont {Becher}, \citenamefont {Roos}, \citenamefont {Eschner},\ and\
  \citenamefont {Blatt}}]{schmidt2003realization}%
  \BibitemOpen
  \bibfield  {author} {\bibinfo {author} {\bibfnamefont {F.}~\bibnamefont
  {Schmidt-Kaler}}, \bibinfo {author} {\bibfnamefont {H.}~\bibnamefont
  {H{\"a}ffner}}, \bibinfo {author} {\bibfnamefont {M.}~\bibnamefont {Riebe}},
  \bibinfo {author} {\bibfnamefont {S.}~\bibnamefont {Gulde}}, \bibinfo
  {author} {\bibfnamefont {G.~P.}\ \bibnamefont {Lancaster}}, \bibinfo {author}
  {\bibfnamefont {T.}~\bibnamefont {Deuschle}}, \bibinfo {author}
  {\bibfnamefont {C.}~\bibnamefont {Becher}}, \bibinfo {author} {\bibfnamefont
  {C.~F.}\ \bibnamefont {Roos}}, \bibinfo {author} {\bibfnamefont
  {J.}~\bibnamefont {Eschner}}, \ and\ \bibinfo {author} {\bibfnamefont
  {R.}~\bibnamefont {Blatt}},\ }\href@noop {} {\bibfield  {journal} {\bibinfo
  {journal} {Nature}\ }\textbf {\bibinfo {volume} {422}},\ \bibinfo {pages}
  {408} (\bibinfo {year} {2003})}\BibitemShut {NoStop}%
\bibitem [{\citenamefont {S{\o}rensen}\ and\ \citenamefont
  {M{\o}lmer}(1999)}]{sorensen1999quantum}%
  \BibitemOpen
  \bibfield  {author} {\bibinfo {author} {\bibfnamefont {A.}~\bibnamefont
  {S{\o}rensen}}\ and\ \bibinfo {author} {\bibfnamefont {K.}~\bibnamefont
  {M{\o}lmer}},\ }\href@noop {} {\bibfield  {journal} {\bibinfo  {journal}
  {Physical review letters}\ }\textbf {\bibinfo {volume} {82}},\ \bibinfo
  {pages} {1971} (\bibinfo {year} {1999})}\BibitemShut {NoStop}%
\bibitem [{\citenamefont {Cabrillo}\ \emph {et~al.}(1999)\citenamefont
  {Cabrillo}, \citenamefont {Cirac}, \citenamefont
  {Garc{\'{\i}}a-Fern{\'{a}}ndez},\ and\ \citenamefont
  {Zoller}}]{cabrillo1999}%
  \BibitemOpen
  \bibfield  {author} {\bibinfo {author} {\bibfnamefont {C.}~\bibnamefont
  {Cabrillo}}, \bibinfo {author} {\bibfnamefont {J.}~\bibnamefont {Cirac}},
  \bibinfo {author} {\bibfnamefont {P.}~\bibnamefont
  {Garc{\'{\i}}a-Fern{\'{a}}ndez}}, \ and\ \bibinfo {author} {\bibfnamefont
  {P.}~\bibnamefont {Zoller}},\ }\href@noop {} {\bibfield  {journal} {\bibinfo
  {journal} {Phys. Rev. A}\ }\textbf {\bibinfo {volume} {59}},\ \bibinfo
  {pages} {1025} (\bibinfo {year} {1999})}\BibitemShut {NoStop}%
\bibitem [{\citenamefont {Slodi{\v{c}}ka}\ \emph {et~al.}(2013)\citenamefont
  {Slodi{\v{c}}ka}, \citenamefont {H{\'{e}}tet}, \citenamefont {R{\"{o}}ck},
  \citenamefont {Schindler}, \citenamefont {Hennrich},\ and\ \citenamefont
  {Blatt}}]{Slodicka2013}%
  \BibitemOpen
  \bibfield  {author} {\bibinfo {author} {\bibfnamefont {L.}~\bibnamefont
  {Slodi{\v{c}}ka}}, \bibinfo {author} {\bibfnamefont {G.}~\bibnamefont
  {H{\'{e}}tet}}, \bibinfo {author} {\bibfnamefont {N.}~\bibnamefont
  {R{\"{o}}ck}}, \bibinfo {author} {\bibfnamefont {P.}~\bibnamefont
  {Schindler}}, \bibinfo {author} {\bibfnamefont {M.}~\bibnamefont {Hennrich}},
  \ and\ \bibinfo {author} {\bibfnamefont {R.}~\bibnamefont {Blatt}},\
  }\href@noop {} {\bibfield  {journal} {\bibinfo  {journal} {Phys. Rev. Lett.}\
  }\textbf {\bibinfo {volume} {110}},\ \bibinfo {pages} {083603} (\bibinfo
  {year} {2013})}\BibitemShut {NoStop}%
\bibitem [{\citenamefont {H{\"a}ffner}\ \emph {et~al.}(2005)\citenamefont
  {H{\"a}ffner}, \citenamefont {Schmidt-Kaler}, \citenamefont {H{\"a}nsel},
  \citenamefont {Roos}, \citenamefont {K{\"o}rber}, \citenamefont {Chwalla},
  \citenamefont {Riebe}, \citenamefont {Benhelm}, \citenamefont {Rapol},
  \citenamefont {Becher},\ and\ \citenamefont {R.}}]{haffner2005robust}%
  \BibitemOpen
  \bibfield  {author} {\bibinfo {author} {\bibfnamefont {H.}~\bibnamefont
  {H{\"a}ffner}}, \bibinfo {author} {\bibfnamefont {F.}~\bibnamefont
  {Schmidt-Kaler}}, \bibinfo {author} {\bibfnamefont {W.}~\bibnamefont
  {H{\"a}nsel}}, \bibinfo {author} {\bibfnamefont {C.}~\bibnamefont {Roos}},
  \bibinfo {author} {\bibfnamefont {T.}~\bibnamefont {K{\"o}rber}}, \bibinfo
  {author} {\bibfnamefont {M.}~\bibnamefont {Chwalla}}, \bibinfo {author}
  {\bibfnamefont {M.}~\bibnamefont {Riebe}}, \bibinfo {author} {\bibfnamefont
  {J.}~\bibnamefont {Benhelm}}, \bibinfo {author} {\bibfnamefont {U.~D.}\
  \bibnamefont {Rapol}}, \bibinfo {author} {\bibfnamefont {C.}~\bibnamefont
  {Becher}}, \ and\ \bibinfo {author} {\bibfnamefont {B.}~\bibnamefont {R.}},\
  }\href@noop {} {\bibfield  {journal} {\bibinfo  {journal} {Appl. Phys. B}\
  }\textbf {\bibinfo {volume} {81}},\ \bibinfo {pages} {151} (\bibinfo {year}
  {2005})}\BibitemShut {NoStop}%
\bibitem [{\citenamefont {Sackett}\ \emph {et~al.}(2000)\citenamefont
  {Sackett}, \citenamefont {Kielpinski}, \citenamefont {King}, \citenamefont
  {Langer}, \citenamefont {Meyer}, \citenamefont {Myatt}, \citenamefont {Rowe},
  \citenamefont {Turchette}, \citenamefont {Itano}, \citenamefont {Wineland},\
  and\ \citenamefont {Monroe}}]{sackett2000experimental}%
  \BibitemOpen
  \bibfield  {author} {\bibinfo {author} {\bibfnamefont {C.~A.}\ \bibnamefont
  {Sackett}}, \bibinfo {author} {\bibfnamefont {D.}~\bibnamefont {Kielpinski}},
  \bibinfo {author} {\bibfnamefont {B.~E.}\ \bibnamefont {King}}, \bibinfo
  {author} {\bibfnamefont {C.}~\bibnamefont {Langer}}, \bibinfo {author}
  {\bibfnamefont {V.}~\bibnamefont {Meyer}}, \bibinfo {author} {\bibfnamefont
  {C.~J.}\ \bibnamefont {Myatt}}, \bibinfo {author} {\bibfnamefont
  {M.}~\bibnamefont {Rowe}}, \bibinfo {author} {\bibfnamefont {Q.~A.}\
  \bibnamefont {Turchette}}, \bibinfo {author} {\bibfnamefont {W.~M.}\
  \bibnamefont {Itano}}, \bibinfo {author} {\bibfnamefont {D.~J.}\ \bibnamefont
  {Wineland}}, \ and\ \bibinfo {author} {\bibfnamefont {C.}~\bibnamefont
  {Monroe}},\ }\href@noop {} {\bibfield  {journal} {\bibinfo  {journal} {Nature
  (London)}\ }\textbf {\bibinfo {volume} {404}},\ \bibinfo {pages} {256}
  (\bibinfo {year} {2000})}\BibitemShut {NoStop}%
\bibitem [{\citenamefont {Suzuki}\ \emph {et~al.}(2010)\citenamefont {Suzuki},
  \citenamefont {Miniatura},\ and\ \citenamefont
  {Nemoto}}]{suzuki2010entanglement}%
  \BibitemOpen
  \bibfield  {author} {\bibinfo {author} {\bibfnamefont {J.}~\bibnamefont
  {Suzuki}}, \bibinfo {author} {\bibfnamefont {C.}~\bibnamefont {Miniatura}}, \
  and\ \bibinfo {author} {\bibfnamefont {K.}~\bibnamefont {Nemoto}},\
  }\href@noop {} {\bibfield  {journal} {\bibinfo  {journal} {Phys. Rev. A}\
  }\textbf {\bibinfo {volume} {81}},\ \bibinfo {pages} {062307} (\bibinfo
  {year} {2010})}\BibitemShut {NoStop}%
\bibitem [{\citenamefont {Scholak}\ \emph {et~al.}(2008)\citenamefont
  {Scholak}, \citenamefont {Mintert},\ and\ \citenamefont
  {M{\"u}ller}}]{scholak2008entanglement}%
  \BibitemOpen
  \bibfield  {author} {\bibinfo {author} {\bibfnamefont {T.}~\bibnamefont
  {Scholak}}, \bibinfo {author} {\bibfnamefont {F.}~\bibnamefont {Mintert}}, \
  and\ \bibinfo {author} {\bibfnamefont {C.~A.}\ \bibnamefont {M{\"u}ller}},\
  }\href@noop {} {\bibfield  {journal} {\bibinfo  {journal} {Europhys. Lett.}\
  }\textbf {\bibinfo {volume} {83}},\ \bibinfo {pages} {60006} (\bibinfo {year}
  {2008})}\BibitemShut {NoStop}%
\bibitem [{\citenamefont {Schmidt-Kaler}\ and\ \citenamefont
  {Gerritsma}(2012)}]{schmidt2012entangled}%
  \BibitemOpen
  \bibfield  {author} {\bibinfo {author} {\bibfnamefont {F.}~\bibnamefont
  {Schmidt-Kaler}}\ and\ \bibinfo {author} {\bibfnamefont {R.}~\bibnamefont
  {Gerritsma}},\ }\href@noop {} {\bibfield  {journal} {\bibinfo  {journal}
  {Europhys. Lett.}\ }\textbf {\bibinfo {volume} {99}},\ \bibinfo {pages}
  {53001} (\bibinfo {year} {2012})}\BibitemShut {NoStop}%
\bibitem [{\citenamefont {Ruster}\ \emph {et~al.}(2017)\citenamefont {Ruster},
  \citenamefont {Kaufmann}, \citenamefont {Luda}, \citenamefont {Kaushal},
  \citenamefont {Schmiegelow}, \citenamefont {Schmidt-Kaler},\ and\
  \citenamefont {Poschinger}}]{ruster2017entanglement}%
  \BibitemOpen
  \bibfield  {author} {\bibinfo {author} {\bibfnamefont {T.}~\bibnamefont
  {Ruster}}, \bibinfo {author} {\bibfnamefont {H.}~\bibnamefont {Kaufmann}},
  \bibinfo {author} {\bibfnamefont {M.}~\bibnamefont {Luda}}, \bibinfo {author}
  {\bibfnamefont {V.}~\bibnamefont {Kaushal}}, \bibinfo {author} {\bibfnamefont
  {C.}~\bibnamefont {Schmiegelow}}, \bibinfo {author} {\bibfnamefont
  {F.}~\bibnamefont {Schmidt-Kaler}}, \ and\ \bibinfo {author} {\bibfnamefont
  {U.}~\bibnamefont {Poschinger}},\ }\href@noop {} {\bibfield  {journal}
  {\bibinfo  {journal} {Physical Review X}\ }\textbf {\bibinfo {volume} {7}},\
  \bibinfo {pages} {031050} (\bibinfo {year} {2017})}\BibitemShut {NoStop}%
\bibitem [{\citenamefont {DeVoe}\ and\ \citenamefont
  {Brewer}(1996)}]{devoe1996observation}%
  \BibitemOpen
  \bibfield  {author} {\bibinfo {author} {\bibfnamefont {R.}~\bibnamefont
  {DeVoe}}\ and\ \bibinfo {author} {\bibfnamefont {R.}~\bibnamefont {Brewer}},\
  }\href@noop {} {\bibfield  {journal} {\bibinfo  {journal} {Phys. Rev. Lett.}\
  }\textbf {\bibinfo {volume} {76}},\ \bibinfo {pages} {2049} (\bibinfo {year}
  {1996})}\BibitemShut {NoStop}%
\bibitem [{\citenamefont {Ficek}\ and\ \citenamefont
  {Tana{\'s}}(2002)}]{ficek2002entangled}%
  \BibitemOpen
  \bibfield  {author} {\bibinfo {author} {\bibfnamefont {Z.}~\bibnamefont
  {Ficek}}\ and\ \bibinfo {author} {\bibfnamefont {R.}~\bibnamefont
  {Tana{\'s}}},\ }\href@noop {} {\bibfield  {journal} {\bibinfo  {journal}
  {Phys. Rep.}\ }\textbf {\bibinfo {volume} {372}},\ \bibinfo {pages} {369}
  (\bibinfo {year} {2002})}\BibitemShut {NoStop}%
\bibitem [{\citenamefont {Navarrete-Benlloch}\ \emph
  {et~al.}(2011)\citenamefont {Navarrete-Benlloch}, \citenamefont {de~Vega},
  \citenamefont {Porras},\ and\ \citenamefont
  {Cirac}}]{navarrete2011simulating}%
  \BibitemOpen
  \bibfield  {author} {\bibinfo {author} {\bibfnamefont {C.}~\bibnamefont
  {Navarrete-Benlloch}}, \bibinfo {author} {\bibfnamefont {I.}~\bibnamefont
  {de~Vega}}, \bibinfo {author} {\bibfnamefont {D.}~\bibnamefont {Porras}}, \
  and\ \bibinfo {author} {\bibfnamefont {J.~I.}\ \bibnamefont {Cirac}},\
  }\href@noop {} {\bibfield  {journal} {\bibinfo  {journal} {New J. Phys.}\
  }\textbf {\bibinfo {volume} {13}},\ \bibinfo {pages} {023024} (\bibinfo
  {year} {2011})}\BibitemShut {NoStop}%
\bibitem [{\citenamefont {Barredo}\ \emph {et~al.}(2016)\citenamefont
  {Barredo}, \citenamefont {de~L{\'e}s{\'e}leuc}, \citenamefont {Lienhard},
  \citenamefont {Lahaye},\ and\ \citenamefont {Browaeys}}]{barredo2016atom}%
  \BibitemOpen
  \bibfield  {author} {\bibinfo {author} {\bibfnamefont {D.}~\bibnamefont
  {Barredo}}, \bibinfo {author} {\bibfnamefont {S.}~\bibnamefont
  {de~L{\'e}s{\'e}leuc}}, \bibinfo {author} {\bibfnamefont {V.}~\bibnamefont
  {Lienhard}}, \bibinfo {author} {\bibfnamefont {T.}~\bibnamefont {Lahaye}}, \
  and\ \bibinfo {author} {\bibfnamefont {A.}~\bibnamefont {Browaeys}},\
  }\href@noop {} {\bibfield  {journal} {\bibinfo  {journal} {Science}\ }\textbf
  {\bibinfo {volume} {354}},\ \bibinfo {pages} {1021} (\bibinfo {year}
  {2016})}\BibitemShut {NoStop}%
\bibitem [{\citenamefont {Jakob}\ and\ \citenamefont
  {Bergou}(2010)}]{jakob2010quantitative}%
  \BibitemOpen
  \bibfield  {author} {\bibinfo {author} {\bibfnamefont {M.}~\bibnamefont
  {Jakob}}\ and\ \bibinfo {author} {\bibfnamefont {J.~A.}\ \bibnamefont
  {Bergou}},\ }\href@noop {} {\bibfield  {journal} {\bibinfo  {journal} {Opt.
  Commun.}\ }\textbf {\bibinfo {volume} {283}},\ \bibinfo {pages} {827}
  (\bibinfo {year} {2010})}\BibitemShut {NoStop}%
\bibitem [{\citenamefont {Vittorini}\ \emph {et~al.}(2014)\citenamefont
  {Vittorini}, \citenamefont {Hucul}, \citenamefont {Inlek}, \citenamefont
  {Crocker},\ and\ \citenamefont {Monroe}}]{vittorini2014entanglement}%
  \BibitemOpen
  \bibfield  {author} {\bibinfo {author} {\bibfnamefont {G.}~\bibnamefont
  {Vittorini}}, \bibinfo {author} {\bibfnamefont {D.}~\bibnamefont {Hucul}},
  \bibinfo {author} {\bibfnamefont {I.~V.}\ \bibnamefont {Inlek}}, \bibinfo
  {author} {\bibfnamefont {C.}~\bibnamefont {Crocker}}, \ and\ \bibinfo
  {author} {\bibfnamefont {C.}~\bibnamefont {Monroe}},\ }\href@noop {}
  {\bibfield  {journal} {\bibinfo  {journal} {Phys. Rev. A}\ }\textbf {\bibinfo
  {volume} {90}},\ \bibinfo {pages} {040302} (\bibinfo {year}
  {2014})}\BibitemShut {NoStop}%
\bibitem [{\citenamefont {Kurizki}\ \emph {et~al.}(2015)\citenamefont
  {Kurizki}, \citenamefont {Bertet}, \citenamefont {Kubo}, \citenamefont {K.},
  \citenamefont {D.}, \citenamefont {Rabl},\ and\ \citenamefont
  {Schmiedmayer}}]{kurizki2015quantum}%
  \BibitemOpen
  \bibfield  {author} {\bibinfo {author} {\bibfnamefont {G.}~\bibnamefont
  {Kurizki}}, \bibinfo {author} {\bibfnamefont {P.}~\bibnamefont {Bertet}},
  \bibinfo {author} {\bibfnamefont {Y.}~\bibnamefont {Kubo}}, \bibinfo {author}
  {\bibfnamefont {M.}~\bibnamefont {K.}}, \bibinfo {author} {\bibfnamefont
  {P.}~\bibnamefont {D.}}, \bibinfo {author} {\bibfnamefont {P.}~\bibnamefont
  {Rabl}}, \ and\ \bibinfo {author} {\bibfnamefont {J.}~\bibnamefont
  {Schmiedmayer}},\ }\href@noop {} {\bibfield  {journal} {\bibinfo  {journal}
  {Proc. Natl. Acad. Sci. U.S.A.}\ }\textbf {\bibinfo {volume} {112}},\
  \bibinfo {pages} {3866} (\bibinfo {year} {2015})}\BibitemShut {NoStop}%
\bibitem [{\citenamefont {Wiegner}\ \emph
  {et~al.}(2011{\natexlab{b}})\citenamefont {Wiegner}, \citenamefont
  {Von~Zanthier},\ and\ \citenamefont {Agarwal}}]{wiegner2011IOP}%
  \BibitemOpen
  \bibfield  {author} {\bibinfo {author} {\bibfnamefont {R.}~\bibnamefont
  {Wiegner}}, \bibinfo {author} {\bibfnamefont {J.}~\bibnamefont
  {Von~Zanthier}}, \ and\ \bibinfo {author} {\bibfnamefont {G.~S.}\
  \bibnamefont {Agarwal}},\ }\href@noop {} {\bibfield  {journal} {\bibinfo
  {journal} {J. Phys. B: At. Mol. Opt. Phys.}\ }\textbf {\bibinfo {volume}
  {44}},\ \bibinfo {pages} {055501} (\bibinfo {year}
  {2011}{\natexlab{b}})}\BibitemShut {NoStop}%
\bibitem [{\citenamefont {Giovannetti}\ \emph {et~al.}(2011)\citenamefont
  {Giovannetti}, \citenamefont {Lloyd},\ and\ \citenamefont
  {Maccone}}]{giovannetti2011advances}%
  \BibitemOpen
  \bibfield  {author} {\bibinfo {author} {\bibfnamefont {V.}~\bibnamefont
  {Giovannetti}}, \bibinfo {author} {\bibfnamefont {S.}~\bibnamefont {Lloyd}},
  \ and\ \bibinfo {author} {\bibfnamefont {L.}~\bibnamefont {Maccone}},\
  }\href@noop {} {\bibfield  {journal} {\bibinfo  {journal} {Nature Photon.}\
  }\textbf {\bibinfo {volume} {5}},\ \bibinfo {pages} {222} (\bibinfo {year}
  {2011})}\BibitemShut {NoStop}%
\bibitem [{\citenamefont {Predehl}\ \emph {et~al.}(2012)\citenamefont
  {Predehl}, \citenamefont {Grosche}, \citenamefont {Raupach}, \citenamefont
  {Droste}, \citenamefont {Terra}, \citenamefont {Alnis}, \citenamefont
  {Legero}, \citenamefont {H{\"a}nsch}, \citenamefont {Udem}, \citenamefont
  {Holzwarth},\ and\ \citenamefont {Schnatz}}]{predehl2012920}%
  \BibitemOpen
  \bibfield  {author} {\bibinfo {author} {\bibfnamefont {K.}~\bibnamefont
  {Predehl}}, \bibinfo {author} {\bibfnamefont {G.}~\bibnamefont {Grosche}},
  \bibinfo {author} {\bibfnamefont {S.~M.~F.}\ \bibnamefont {Raupach}},
  \bibinfo {author} {\bibfnamefont {S.}~\bibnamefont {Droste}}, \bibinfo
  {author} {\bibfnamefont {O.}~\bibnamefont {Terra}}, \bibinfo {author}
  {\bibfnamefont {J.}~\bibnamefont {Alnis}}, \bibinfo {author} {\bibfnamefont
  {T.}~\bibnamefont {Legero}}, \bibinfo {author} {\bibfnamefont {T.~W.}\
  \bibnamefont {H{\"a}nsch}}, \bibinfo {author} {\bibfnamefont
  {T.}~\bibnamefont {Udem}}, \bibinfo {author} {\bibfnamefont {R.}~\bibnamefont
  {Holzwarth}}, \ and\ \bibinfo {author} {\bibfnamefont {H.}~\bibnamefont
  {Schnatz}},\ }\href@noop {} {\bibfield  {journal} {\bibinfo  {journal}
  {Science}\ }\textbf {\bibinfo {volume} {336}},\ \bibinfo {pages} {441}
  (\bibinfo {year} {2012})}\BibitemShut {NoStop}%
\bibitem [{\citenamefont {Yurtsever}\ \emph {et~al.}(2003)\citenamefont
  {Yurtsever}, \citenamefont {Strekalov},\ and\ \citenamefont
  {Dowling}}]{yurtsever2003interferometry}%
  \BibitemOpen
  \bibfield  {author} {\bibinfo {author} {\bibfnamefont {U.}~\bibnamefont
  {Yurtsever}}, \bibinfo {author} {\bibfnamefont {D.}~\bibnamefont
  {Strekalov}}, \ and\ \bibinfo {author} {\bibfnamefont {J.~P.}\ \bibnamefont
  {Dowling}},\ }\href@noop {} {\bibfield  {journal} {\bibinfo  {journal} {Eur.
  Phys. J. D}\ }\textbf {\bibinfo {volume} {22}},\ \bibinfo {pages} {365}
  (\bibinfo {year} {2003})}\BibitemShut {NoStop}%
\end{thebibliography}

%

\end{document}


\setlength\linenumbersep{8 mm}	
\title{Supplementary information for \\Interference of single photons emitted by\\entangled atoms in free space}
\author{G. Araneda$^{1}$}
\email{gabriel.araneda-machuca@uibk.ac.at}
\author{D. B. Higginbottom$^{1,2}$}
\author{L. Slodi\v{c}ka$^{3}$}
\author{Y. Colombe$^1$}
\email{yves.colombe@uibk.ac.at}
\author{R. Blatt$^{1,4}$}
\affiliation{$^1$Institut f\"{u}r Experimentalphysik, Universit\"{a}t Innsbruck, Technikerstra\ss e 25, 6020 Innsbruck, Austria}
\affiliation{$^2$Centre for Quantum Computation and Communication Technology, Research School of Physics and Engineering, The Australian National University, Canberra ACT 2601, Australia.}
\affiliation{$^3$Department of Optics Palack\'{y} University, 17. Listopadu 12, 77146 Olomouc, Czech Republic}
\affiliation{$^4$Institut f\"{u}r Quantenoptik und Quanteninformation, \"{O}sterreichische Akademie der Wissenschaften, Technikerstra\ss e 21a, 6020 Innsbruck, Austria}

\maketitle

\renewcommand{\figurename}{Supplementary Figure}
\renewcommand{\thefigure}{S\arabic{figure}}
\setcounter{figure}{0} 
\renewcommand\thesubsection{\arabic{subsection}}

\subsection{Entanglement generation}
\label{SI_EntanglementGeneration}
The two $^{138}$Ba$^{+}$ ions are confined in a linear Paul trap and Doppler-cooled using the $6\text{S}_{1/2} - 6\text{P}_{1/2}$ transition driven by a $\lambda = 493$\,nm laser beam together with a co-propagating repump beam at 650\,nm driving the $5\text{D}_{3/2} - 6\text{P}_{1/2}$ transition \cite{slodivcka2012interferometric}. These two beams are oriented at $45^{\circ}$ with respect to the trap axis to simultaneously cool the radial and axial motional modes. A magnetic field $B = 0.453$\,mT parallel to the ion crystal axis creates a Zeeman splitting of 12.7\,MHz between the ground states $|\text{g}_-\rangle = |6\text{S}_{1/2}, m_j = -\frac{1}{2}\rangle$ and $|\text{g}_+\rangle = |6\text{S}_{1/2}, m_j = +\frac{1}{2}\rangle$.

\begin{figure}[b]
\centerline{\includegraphics[width=0.95\columnwidth]{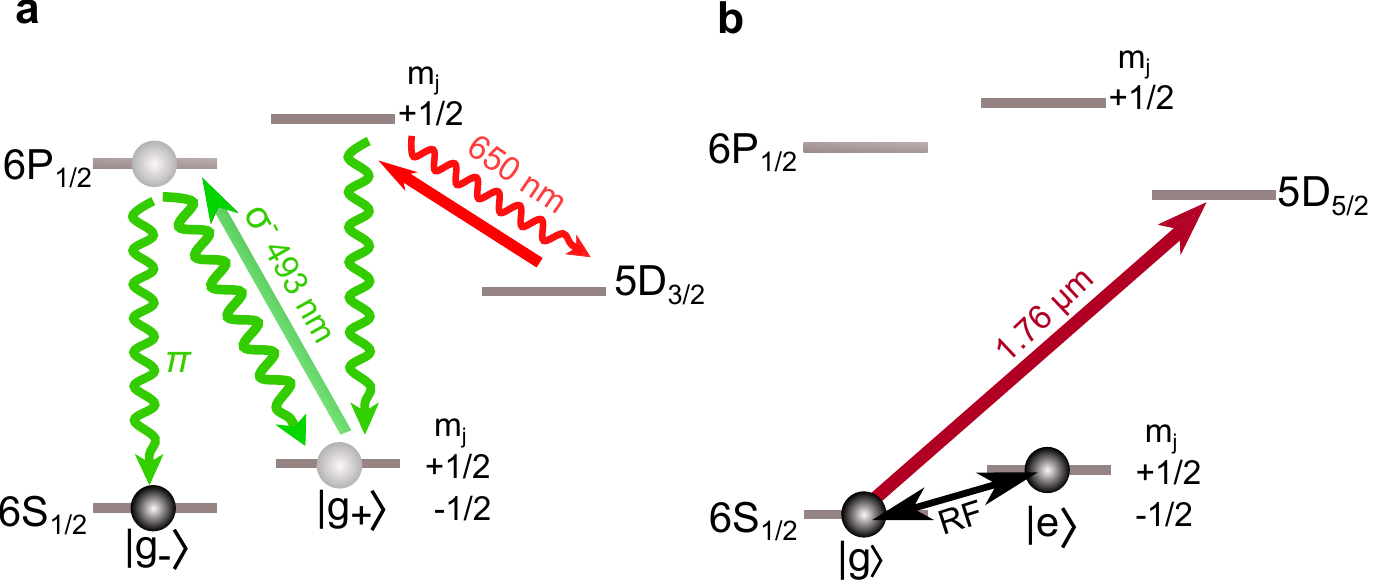}}
\caption{\textbf{Electronic transitions for optical pumping and electron shelving.} \textbf{a}, Both ions are prepared in the initial state $|\text{g}\rangle$ by optical pumping with $\sigma^-$-polarized 493\,nm light and a 650\,nm repump beam. \textbf{b}, RF pulses drive global qubit rotations for parity measurements. Populations in $|\text{g}_-\rangle$ and $|\text{g}_+\rangle$ are determined by a fluorescence measurement after shelving $|\text{g}_-\rangle$ to the metastable state $5\text{D} _{5/2}$ with a narrowband infrared laser.}
\label{atom_structure}
\end{figure}

Both ions are prepared in the state $|\text{g}_-\rangle$ by optical pumping with a $\sigma^-$-polarized 493\,nm laser beam propagating parallel to the magnetic field (Supplementary Fig.~\ref{atom_structure}a). In order to create entanglement we weakly drive the Raman process $|\text{g}_-\rangle\rightarrow|\text{i}\rangle\rightarrow|\text{g}_+\rangle$ as described in the main text, emitting a $\pi$-polarized photon. The collected emission of both ions is superimposed into the single mode fiber (SMF) using the in-vacuum $\text{NA} = 0.40$ objectives L1 and L2 and a distant mirror M (Fig.~1c, main text).

The Raman process is implemented by applying a weak global $\sigma^{+}$-polarized exciting laser pulse along the ion crystal, with $48$\,ns length and $95$\,nW mean power. The strength of the excitation pulse is set in order to obtain a single-ion $|\text{g}_-\rangle\rightarrow|\text{g}_+\rangle$ transition probability $p_\text{e} = 6 \pm 1\,\%$, with a probability $2p_\text{e}(1-p_\text{e})\eta$ of detecting an entanglement heralding event. The value chosen for $p_\text{e}$ corresponds to a trade-off between entanglement generation rate and suppression of double excitations \cite{cabrillo1999}.

Indistinguishability of the detected single photons emitted by the two atoms is essential for the creation of entanglement. While the PBS and the SMF ensure that the detected photons have the same polarization and spatial mode, the indistinguishability is reduced by a detection rate from atom A amounting to only 60\% of that from atom B. This imbalance is due to larger optical losses for the atom A due to additional propagation through an in-vacuum lens, optical viewport, and reflection on the distant mirror, and to spatial mode mismatch after propagation along this path. This imbalance is compensated by a  precise angular adjustment of the distant mirror and of the SMF coupler, reducing the detection rate of atom B to that of atom A.
 
The degree of indistinguishability is characterized prior to the entanglement generation experiment by measuring the second-order correlation function $g^{(2)}(t)$ of the fluorescence light emitted by the two atoms in the common mode using a Hanbury Brown and Twiss detection scheme (see Supplemental material in \cite{Slodicka2013}). For this, the $\pi$-polarized output of the PBS is coupled to a single-mode fiber 50/50 beam splitter, and detected with an APD at each output. The measured indistinguishability is $g^{(2)}(t = 0) = 0.99 \pm 0.06$, close to the ideal value of 1 for completely indistinguishable emitters. Totally distinguishable photons would give $g^{(2)}(t = 0) = 0.5$ \cite{Slodicka2013}.

\begin{figure}[t!]
\centerline{\includegraphics[width=0.95\columnwidth]{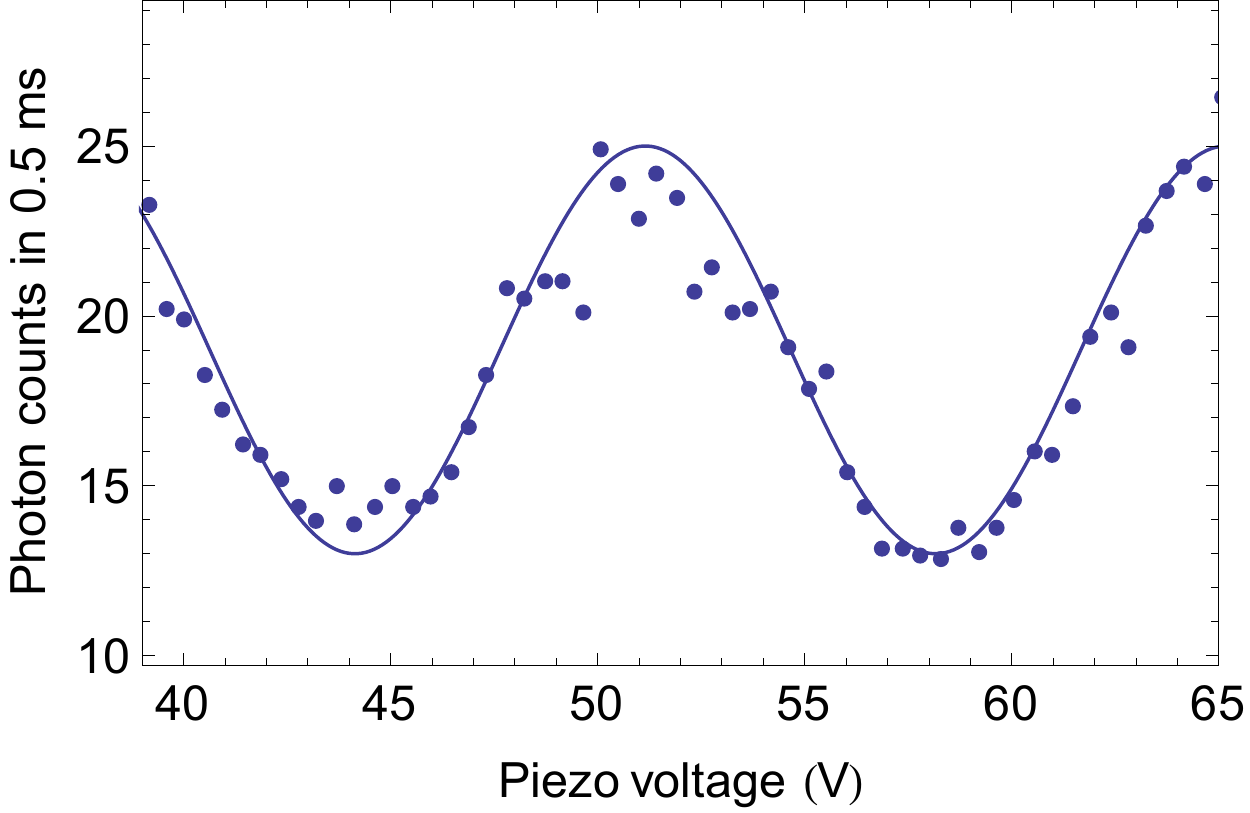}}
\caption{\textbf{Interference in the fluorescence signal of two trapped ions.} $\sigma^+$ and $\sigma^{-}$ photons scattered by the trapped ions during the cooling stage are detected while changing the position of the distant mirror with a slow voltage ramp of 1\,V.s$^{-1}$ on the PZT transducers. The observed interference signal has a visibility $0.37 \pm 0.02$. The PZT voltage corresponding to a full fringe (mirror displacement equal to $\lambda /2$) is $\Delta U = 14.0\,$V.}
\label{interference}
\end{figure}

The phase $\phi$ of the entangled state $|\psi\rangle = \frac{1}{\sqrt{2}}(|\text{g}_+,\text{g}_-\rangle + e^{i\phi}|\text{g}_-,\text{g}_+\rangle)$ is the sum of the phase difference $\phi_{\text{L}}$ in the laser beam driving the two atoms and the phase difference $\phi_{\text{D}}$ associated with the two detection path lengths. The phase $\phi_{\text{L}}$ is given by the atoms separation $z$ along the path of the driving beam $\phi_{\text{L}} = \phi_{\text{L}_\text{B}}-\phi_{\text{L}_\text{A}} = k z$ where $k$ is the wavevector of the driving beam $k = 2\pi / \lambda$, while $\phi_{\text{D}} = \phi_{\text{D}_\text{B}}-\phi_{\text{D}_\text{A}} = - k d$, where $d$ is twice the optical path between the ion crystal and the distant mirror.
The phase of the entangled state is therefore $\phi = k (z-d)$, and can be adjusted by varying the mirror position using three piezoelectric transducers supplied with the same voltage, placed on the mount of the mirror. After recording the interference fringes of the $\sigma^+$ and $\sigma^{-}$-polarized photons scattered during the cooling stage, detected trough coupling the second output of the PBS to a single-mode fiber and a second APD (Supplementary Fig.~\ref{interference}), one can set $\phi$ to the desired value. Note that the phase $\phi$ of the entangled state only needs to be set when characterizing the entangled state through parity oscillations (see Supplementary Section \ref{SI_EntanglementCharacterization}). The scattering probability of a witness photon from the entangled state $|\psi\rangle$ depends solely on the phase \emph{difference} $\Delta\phi = \phi- \phi'$ between the herald and witness detections (see Supplementary Section \ref{SI_FastDisplacement}).

\subsection{Entanglement characterization}
\label{SI_EntanglementCharacterization}
We characterize the entangled state following the procedure described in \cite{Slodicka2013}. After the detection of a herald photon the populations of $\rho_{g_-,g_-}$, $\rho_{g_+,g_+}$ and $\rho_{g_+,g_-}\,+\,\rho_{g_-,g_+}$ of the experimentally generated state $\rho$ are measured by electron shelving; the $|\text{g}_-\rangle$ population is first transferred to the metastable state $|5\text{D}_{5/2},m_j = -\frac{5}{2}\rangle$ with a global pulse of the narrowband $1.76\,\mu$m laser (Supplementary Fig.~\ref{atom_structure}b) before fluorescence is measured by excitation with the 493\,nm cooling beam. The strength of the fluorescence allows to distinguish between states with zero, one and two excitations. From the population measurement, $91 \pm 3$\,\% of the trigger events correspond to a single excitation, while the remaining 9\,\% correspond to zero and double excitation.

Measuring the oscillations in the expectation value of the parity operator \cite{sackett2000experimental} after driving global rotations in the qubit basis $\{|\text{g}_-\rangle,|\text{g}_+\rangle\}$ with an RF field at 12.7 MHz (produced by an external antenna) allows us to confirm the presence of entanglement and bound the coherence between $|\text{g}_+,\text{g}_-\rangle$ and $|\text{g}_-,\text{g}_+\rangle$. Assuming that all other coherences are zero as there is no mechanism that would generate them, we estimate a fidelity of the state generated with $\phi = 0$ with the entangled state $|\psi\rangle$ (Eq.~1 in main text) of $F = 0.65 \pm 0.02$. We observe that the parity operator oscillates with amplitude $1.2 \pm 0.1$, similar to the previous realization \cite{Slodicka2013}. The fidelity is limited largely by atomic motion. \now{After Doppler cooling, the amplitude of the atomic motion in the trap is $\approx 30$ nm in both radial directions. This motion, which is a superposition of the in-phase and out-of-phase modes, produces a fundamental uncertainty in the atomic positions at the moment of the emission, and a corresponding phase uncertainty in the emitted photon. A mathematical description of the effect is found in the Supplementary Information of \cite{Slodicka2013}. The effect of the motion can be reduced, e.g., by performing sideband cooling to the ground state of motion. However, experimental sequences including ground state cooling have a duration nearly ten times that of sequences with Doppler cooling only, lowering the entanglement rate dramatically and making them impractical.} Other effects reducing the fidelity are the atomic recoil during Raman scattering, initial population imperfections and qubit decoherence (the system is taken out of the decoherence-free subspace for parity measurements, with a coherence time $\sim 120\,\mu$s).

With the same assumptions it is possible to estimate the concurrence of the experimentally created entangled state, $\mathcal{C}(\rho)\approx 0.31$. Using the general inequality for density matrices $\rho_{n,n} \rho_{m,m}\le |\rho_{n,m}|^2$ it can be shown that the presence of coherences neglected in our approximation would not decrease the value of the concurrence.

\subsection{Fast tuning of the optical path between the atoms}
\label{SI_FastDisplacement}
\begin{figure}[ht!]
\centerline{\includegraphics[width=0.95\columnwidth]{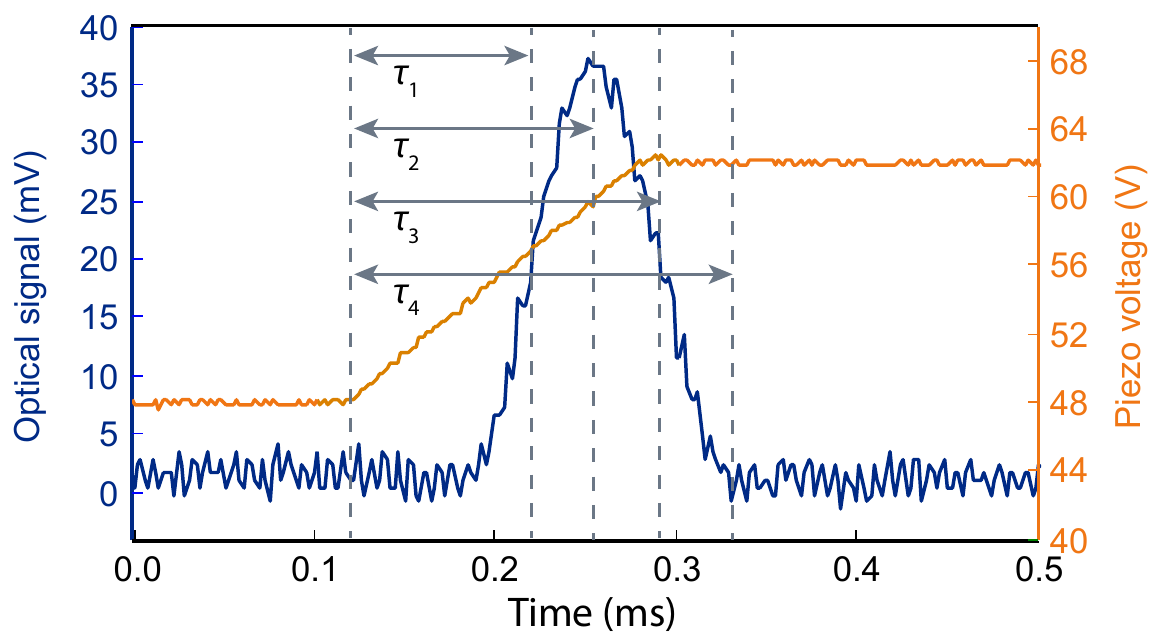}}
\caption{\textbf{Distant mirror displacement calibration.} The distant mirror is displaced by the length corresponding to a full interference fringe by applying a fixed $\simeq 200\,\mu$s linear voltage ramp of 14.0\,V (orange trace) to the PZT transducers. The displacement of the mirror is measured with a Michelson interferometer attached to the set-up. The blue trace shows the recorded interferometric signal. Delay times $\tau_i$ corresponding to different optical phases changes $\Delta \phi_i$ are shown.}
\label{mirror_delay_2}
\end{figure}
In order to vary the phase difference $\Delta\phi = \phi- \phi'$ and observe the interference fringes in the single photon emission from the entangled pair of atoms $|\psi\rangle$, the optical path between the atoms in the common mode is rapidly changed from $d$ to $d'$ between the emissions of the herald and witness photons. This is achieved by displacing the distant mirror using its piezoelectric transducers while keeping the distance $z$ of the ions in the trap constant, resulting in a phase difference $\Delta\phi = k(d'-d)$. After detecting a heralding photon, a linear voltage ramp with amplitude $\Delta U = 14.0\,$V and duration $\Delta T = 200\,\mu$s is applied to the PZTs, displacing the mirror by $(d'-d)/2 = \lambda/2$, corresponding to $\Delta\phi = 2 \pi$ (a full interference fringe). After a variable delay $\tau$ relative to the start of the ramp, an APD detection window of $1\,\mu$s is opened and a 250\,ns, $15.8\,\mu$W laser pulse is applied to the atom pair, which may emit a witness photon. The voltage $\Delta U$ is obtained from the interference signal of the laser-cooled atoms (see Supplementary Section \ref{SI_EntanglementGeneration} and Supplementary Fig.~\ref{interference} above). For the calibration of the phase change $\Delta\phi$ as a function of the delay $\tau$ we implement a simple method using the fact that the photon emission from the entangled atom pair depends only on the \emph{relative} phase $\Delta\phi$. A Michelson interferometer is inserted into the experiment using a flip mirror (Fig.~1c in the main text). The signal of the interferometer is recorded at $\lambda = 493\,$nm for the voltage ramp described above, with an offset voltage of 47.9\,V (Supplementary Fig.~\ref{mirror_delay_2}). Some of the delay times used in the experiment are $\tau_0 = 60\,\mu$s for $\Delta \phi = 0$, $\tau_1 = 93\,\mu$s for $\Delta \phi = \pi/2$, $\tau_2 = 129\,\mu$s for $\Delta \phi = \pi$, $\tau_3 = 164\,\mu$s for $\Delta \phi = 3\pi/2$ and $\tau_4 = 203\,\mu$s for $\Delta \phi = 2\pi$. To obtain the phase difference $\Delta \phi = 5\pi/2$ we apply a larger voltage step of  18.0\,V and a delay time of $220\,\mu$s. The angular alignment of the distant mirror is adjusted every 1\,h by maximizing the visibility of the interference fringes shown in Supplementary Fig.~\ref{interference}.

\subsection{Full experimental sequence}
Each sequence begins with $300\,\mu$s of Doppler cooling followed by up to 30 attempts to generate entanglement. Each of these attempts consists of $20\,\mu$s of optical pumping with a $\sigma^{-}$-polarized pulse to initialize the state $|\text{g}_-,\text{g}_-\rangle$, followed by a weak $\sigma^{+}$-polarized excitation pulse with 48\,ns length and 95\,nW mean power, such that the probability of the Raman process $|\text{g}_-\rangle\rightarrow|\text{g}_+\rangle$ is $p_{\text{e}} = 6 \pm 1$\% for a single atom. The second part of the sequence is triggered by the detection of a photon in the common optical mode by APD, heralding entanglement.
Photons emitted in the other possible decay channel, with wavelength 650\,nm, are filtered out using a bandpass filter with 97\,\% transmittance at 493\,nm and 20\,nm bandwidth. The overall detection efficiency is $\eta \simeq 0.002$, constrained primarily by the $\text{NA} = 0.40$ of the in-vacuum lenses, which allows us to collect $\sim 6\,\%$ of the spontaneous emission from the desired transition in the common mode. Additional optical losses are caused by the lenses and viewports, reflection from the distant mirror, the rest of the optical elements and imperfect coupling into the single-mode fiber, with combined transmission $T \simeq 0.07$. The quantum efficiency of the APD is $\sim 70\,$\%.

During the second part of the sequence the mirror is displaced and a witness photon is generated by a $\sigma^+$-polarized pulse with 250\,ns length and 15.8\,$\mu$W mean power, such that $p_{\text{w}} = 80 \pm 2$\,\% for a single atom. This pulse can be chosen substantially stronger than the entanglement generation pulse because the atom pair in state $|\psi\rangle$ cannot emit multiple photons. The sequence is repeated at a rate 704\,s$^{-1}$, creating $5.02$ entanglement events per second, while the witness photon is detected at an average rate 0.47\,min$^{-1}$. Due to software and hardware improvements in the experiment control system we have been able to generate entanglement at a rate 19 times higher than in our previous realization \cite{Slodicka2013} without reducing the fidelity of the generated entangled state.

\subsection{Emission from entangled and separable states}
\label{SI_EmissionTheory}

\begin{figure}[b]
\centerline{\includegraphics[width=0.95\columnwidth]{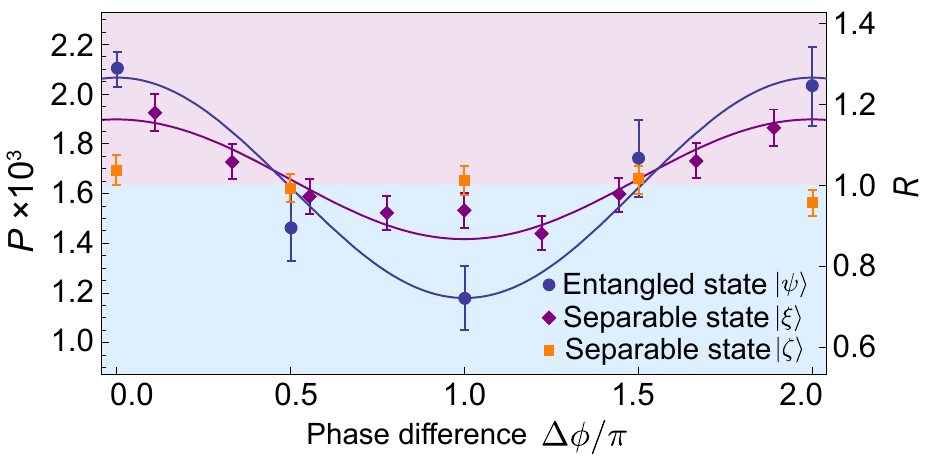}}
\caption{\textbf{Probability of single photon detection for different states}. The measured probability of single photon detection from the separable state $|\xi\rangle$ is shown with purple diamonds. The fitted amplitude corresponds to a visibility $V_{|\eta\rangle}= 0.15 \pm 0.08$. For this case, the mean distance of the mirror was reduced to $\sim 15\,$cm. The measured data for the entangled state $|\psi\rangle$ (blue circles) and for the separable state $|\zeta\rangle$ (orange squares) presented in Fig.~2 (main text) are shown for reference.}
\label{eta}
\end{figure}

Following the calculations in \cite{wiegner2011a} and considering that in our case a $\pi$-polarized photon is emitted every time the transition $|g_-\rangle \rightarrow |g_+\rangle$ is completed, the probability of detecting a photon from a pair of atoms in the common mode is proportional to the intensity
\begin{align}
I= \sum_{i,j}\langle\hat{s}^+_i \hat{s}^-_j\rangle e^{i(\varphi_i-\varphi_j)}\label{eq1}
\end{align}
where $i,j$ denote the atoms A or B, $\hat{s}^-_i=|g_+\rangle_i\langle g_-|_i$ is the effective dipole operator, $\hat{s}^+_i=\hat{s}^{-\dagger}_i$, and $\varphi_i = \phi_{\text{L}_i}+\phi_{\text{D}_i}$.

Let us first consider states with a single excitation, \emph{i.e.}, belonging to the subspace defined by the basis $\{|g_-,g_+\rangle,|g_+,g_-\rangle\}$.
From Eq.~(\ref{eq1}), the intensity for the entangled state $|\psi\rangle$, $I_{|\psi\rangle}=1+\cos(\Delta\phi)$, shows a cosine dependence, while for the separable state $|\zeta\rangle$, $I_{|\zeta\rangle}=1$ is constant. The visibility of the intensity $I$, defined as $(I_{\text{max}}-I_{\text{min}})/(I_{\text{max}}+I_{\text{min}})$ are $V_{|\psi\rangle}=1$ and $V_{|\zeta\rangle}=0$ for $|\psi\rangle$ and $|\zeta\rangle$ respectively. For a general state in this subspace $|\alpha\rangle=a|\text{g}_-,\text{g}_+,\rangle+b|\text{g}_+,\text{g}_-,\rangle$ the visibility is $V_{|\alpha\rangle}=2|ab|$, which equals the concurrence of the state, $\mathcal{C}_{|\alpha\rangle}=V_{|\alpha\rangle}$. The measured probabilities for $|\psi\rangle$ and $|\zeta\rangle$ are presented in Fig.~2 in the main text, showing agreement between the measured visibility and the expected concurrence.

For the most general state of the bipartite system $|\beta\rangle=a|\text{g}_-,\text{g}_+,\rangle+b|\text{g}_+,\text{g}_-,\rangle+c|\text{g}_-,\text{g}_-,\rangle+d|\text{g}_+,\text{g}_+\rangle$, the intensity is
\begin{align}
I_{|\beta\rangle}=|a|^2+|b|^2+2|c|^2+2\Re(a^*b e^{i(\varphi_A-\varphi_B)}).
\end{align}
The visibility is
\begin{align}
V_{|\beta\rangle}=\frac{2|ab|}{|a|^2+|b|^2+2|c|^2},
\label{eq3}
\end{align}
while the concurrence is given by $\mathcal{C}_{|\beta\rangle}=2|cd-ab|$. 
Eq.~(\ref{eq3}) shows that a separable state can still lead to interference, so that, in general, the visibility does not correspond to the concurrence. For the separable state $|\xi\rangle=\frac{1}{2}(|g_-\rangle +|g_+\rangle)\otimes(|g_-\rangle+|g_+\rangle)$, where $a=b=c=d=1/2$, the visibility is $V_{|\xi\rangle}=\frac{1}{2}$, while the concurrence is $\mathcal{C}_{|\eta\rangle} = 0$. The single photon detection probability for the state $|\xi\rangle$, prepared using RF global pulses on the $|g_-\rangle \rightarrow |g_+\rangle$ transition, is shown in Supplementary Fig.~\ref{eta}. We observe a visibility $V_{|\xi\rangle}= 0.15 \pm 0.08$, which is approximately, as expected, half the visibility $V_{|\psi\rangle}= \mathcal{C}_{\text{wit}} = 0.27 \pm 0.03$ measured for the entangled state (see main text).

%